\documentclass[
reprint, 
amsfonts, 
amssymb, amsmath, aps, prl, 
floatfix, twoside, 
superscriptaddress,
bibnotes,
]{revtex4-1}

\usepackage{graphicx}
\usepackage{dcolumn}
\usepackage[english]{babel}
\usepackage[utf8]{inputenc}
\usepackage{amsthm}
\usepackage{mathtools}
\usepackage{physics}
\usepackage{xcolor,colortbl}
\usepackage{adjustbox}
\usepackage{placeins}
\usepackage[T1]{fontenc}
\usepackage{lipsum}
\usepackage{csquotes}
\usepackage{hyperref}
\usepackage{comment}
\usepackage[bb=boondox]{mathalfa}
\usepackage[normalem]{ulem}
\usepackage{diagbox}
\usepackage{slashbox}
\usepackage{tabularx,booktabs}
\usepackage{MnSymbol}

\newcommand{\aop}{\hat a}
\newcommand{\adagop}{\hat a^{\dagger}}

\newcommand{\phir}{\hat\phi_{r}}
\newcommand{\phit}{\hat\phi_{t}}
\newcommand{\nr}{\hat n_{r}}
\newcommand{\nt}{\hat n_{t}}

\begin{document}

\title{Readout-Induced Leakage in Superconducting Circuits with Nonlinear Couplings}

%authorlist
\author{Sumeru Hazra}
\thanks{These three authors contributed equally: sumeru.hazra@kit.edu, wei.dai@quantum-machines.co, daniel.kamrath.weiss@gmail.com}
\author{Wei Dai}
\thanks{These three authors contributed equally: sumeru.hazra@kit.edu, wei.dai@quantum-machines.co, daniel.kamrath.weiss@gmail.com}
\affiliation{Department of Applied Physics, Yale University, New Haven, CT 06520, USA}
\author{Daniel K. Weiss}
\thanks{These three authors contributed equally: sumeru.hazra@kit.edu, wei.dai@quantum-machines.co, daniel.kamrath.weiss@gmail.com}
\affiliation{Department of Applied Physics, Yale University, New Haven, CT 06520, USA}
\affiliation{Yale Quantum Institute, Yale University, New Haven, Connecticut 06520, USA}
\author{Pranav D. Parakh}
\affiliation{Department of Applied Physics, Yale University, New Haven, CT 06520, USA}
\author{Luigi Frunzio}
\affiliation{Department of Applied Physics, Yale University, New Haven, CT 06520, USA}
\affiliation{Yale Quantum Institute, Yale University, New Haven, Connecticut 06520, USA}
\author{Michel H. Devoret}
\email{devoret@ucsb.edu}
\affiliation{Department of Applied Physics, Yale University, New Haven, CT 06520, USA}
\affiliation{Department of Applied Physics, University of California Santa Barbara, California 93106, USA}
\affiliation{Google Quantum AI, Santa Barbara, California, USA}
\date{\today}

\begin{abstract}
In superconducting circuits, drive-induced unwanted transitions limit the readout power, thereby constraining readout speed and fidelity. 
When such transitions excite the qubit into leakage states, they produce correlated errors that are particularly harmful for quantum error correction. 
Native nonlinear qubit-readout resonator coupling is a promising alternative to conventional linear hybridization because it provides intrinsic Purcell protection and stricter selection rules for multiphoton processes.
In realistic devices, however, we show that such a coupling alone neither eliminates nor necessarily suppresses drive-induced transitions.
Instead, if not appropriately engineered, these couplings often worsen the situation by introducing additional parasitic processes. 
Moreover, the rates of these unwanted transitions remain sensitive to the choice of readout frequency, regardless of the coupling mechanism. We demonstrate that readout-induced leakage can thus vary by orders of magnitude even when readout frequencies differ by less than $~7\%$.
Our results establish that the benefits of native nonlinear couplings are realized only through informed device design, including the spectral placement of relevant auxiliary modes and elimination of parasitic ones. 
\end{abstract}

\maketitle
Realizing large-scale fault-tolerant quantum computation requires the precise implementation of a set of fundamental operations~\cite{divincenzo2000}, including state preparation~\cite{riste2012_reset, Zhou2021RapidReset}, gate operations~\cite{ding2023_fluxonium_gate,li2024_cz}, and qubit readout~\cite{swiadek2024, spring2025, kurilovich2025}.
Among these operations, achieving fast, high-fidelity quantum non-demolition readout has emerged as a key challenge in state-of-the-art quantum processors~\cite{google2025quantum}, directly impacting the performance of quantum error correction. In dispersive readout~\cite{gambetta_dispersive}, increasing drive power increases signal-to-noise ratio and reduces the readout duration, thereby increasing readout fidelity for a given decoherence time of the qubit. In practice, however, this strategy encounters a fundamental obstacle: beyond a certain drive strength, the readout operation often excites the qubit into the non-computational `leakage' states~\cite{Sank_2016_MIST, Khezri_2023_MIST, verney_2019_ist}.
Such leakage processes generate correlated errors and pose a major challenge for fault-tolerant operation.
Readout-induced leakage arises from multiphoton resonances~\cite{Sank_2016_MIST, Khezri_2023_MIST, Dumas_2024_Ionization, xiao_2025_diagram,kurilovich2025,dai_hazra_weiss_2025, benhayounekhadraoui2025}, in which an off-resonant measurement drive excites the physical qubit into some highly-excited state through the absorption of multiple photons. 
These resonances, in general, connect any pair of states in the composite Hilbert space of the qubit and any additional modes which couple to the qubit~\cite{dai_hazra_weiss_2025, connolly_2025}. 

\begin{figure}[t]
\includegraphics[width = \columnwidth]{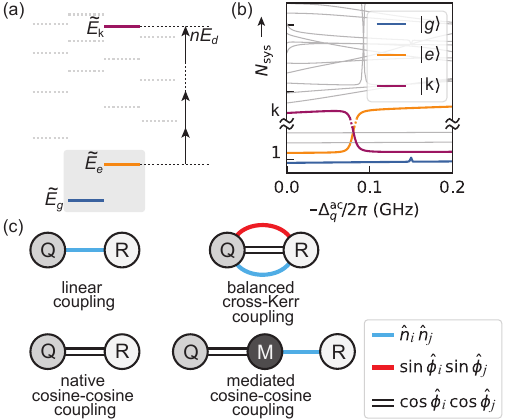}
\caption{Drive-induced leakage in superconducting circuits. 
(a) Dressed energy-level diagram of a multi-mode system. A drive-induced unwanted state transition (DUST), e.g. $|e\rangle\rightarrow |k\rangle$, is activated when a multi-quanta resonance condition is met. 
(b) Floquet branch analysis of the driven system at a given drive frequency, showing a branch swap between the dressed $|e\rangle$ and $|k\rangle$ levels when the drive power (parameterized by the induced ac-Stark shift) meets a multi-quanta resonance condition. (c) Schematic of different qubit (Q)-readout (R) coupling mechanisms $\hat O_i\hat O_j, \{i, j\}\in\{q, m, r\}$: %perturbative cross-Kerr 
linear coupling via $\hat n_q\hat n_r$ interaction (blue line); balanced cross-Kerr coupling via a combination of $\hat n_q\hat n_r$, $\cos\hat \phi_q\cos\hat \phi_r$, and $\sin\hat \phi_q\sin\hat \phi_r$ interactions (red line); a native cosine-cosine, i.e., $\cos\hat \phi_q\cos\hat \phi_r$ interaction (double solid line); and a mediated cosine-cosine coupling via an auxiliary (mediator) mode, leading to  $\cos\hat \phi_q\cos (\hat\phi_m+\hat \phi_r) $ interaction.
}
\label{fig:intro} 
\end{figure}

As illustrated in Fig.~\ref{fig:intro}(a), a transition from $\ket{e}$ to $\ket{k}$ occurs (barring forbidden transitions) when the energy difference between these two driven or \textit{dressed} states satisfies $\tilde E_k-\tilde E_e=nE_d, n\in \mathbb N$, where $E_d$ is the energy of the drive photon, and $\tilde E_i$ represents the energy of the dressed-driven states. 
Since different energy levels acquire different ac-Stark shifts, the resonance conditions evolve with drive power. 
Also, at higher power, the resonance broadens, and the transition strength increases, making the qubit more susceptible to leakage. Additional fluctuations, such as offset charge drift~\cite{dai_hazra_weiss_2025, Fechant2025}, further spread the resonance conditions. 
As a consequence, the qubit generally encounters leakage channels as the drive power is increased. 
The onset of such processes is usually marked by a ``critical drive power'', which is predicted within a Floquet description~\cite{verney_2019_ist, Dumas_2024_Ionization, xiao_2025_diagram}. In this simulation framework, multiphoton resonances appear as branch swaps in the spectrum dressed by the drive. This swap corresponds to hybridization between states~\cite{Dumas_2024_Ionization}. 
Figure~\ref{fig:intro}(b) shows such a branch swap between $\ket{e}$ and $\ket{k}$, which will result in leakage from the computational subspace.

While multiphoton resonances are intrinsic to nonlinear circuits, symmetry-induced or engineered selection rules suppress many of these resonances~\cite{yao_aniket_2023, chapple2024_long, xiao_2025_diagram}. 
This principle motivates several engineered \textit{nonlinear} coupling schemes\cite{didier_2015, gambetta_2011_dimon, dassonneville_2020, potts_2025_long}\footnote{All these nonlinear coupling schemes intrinsically eliminate Purcell decay of the qubit mode.}. These couplings can be broadly categorized into the following types: balanced cross-Kerr coupling~\cite{bright2024, chapple2025_bal, wang_2025_bal_exp, beaulieu2026_bal_exp}, 
native cosine-cosine coupling~\cite{didier_2015,chapple2024_long, salunkhe_2025}, and mediated cosine-cosine coupling~\cite{hazra_dai_2025, mori2025_nonlin, mori2025}, illustrated schematically in Fig.~\ref{fig:intro}(c).
In conventional architectures, charge-charge coupling (blue line) between the transmon~\cite{transmon_2007} (Q) and the readout resonator (R)  linearly hybridizes the modes and perturbatively generates an effective cross-Kerr interaction. Engineered approaches, such as balanced cross-Kerr coupling, modify this interaction by combining charge-charge coupling with a junction coupling, producing sine-sine (red line) and cosine-cosine (double black line) interactions to cancel the linear exchange term. In the ideal limit, a native cosine-cosine interaction eliminates several multiphoton transitions through selection rules. However, in realistic implementations, this interaction is typically mediated by auxiliary modes (M)~\cite{dassonneville_2020, mori2025, hazra_dai_2025} or inevitably accompanied by additional modes arising from parasitic capacitance~\cite{salunkhe_2025}. What remains of the protection offered by \textit{nonlinear coupling schemes} once these \textit{additional modes} are included?

In this Letter, we numerically analyze the three nonlinear coupling schemes under idealized symmetry conditions and benchmark them against a linearly coupled transmon-resonator system.
We show that, among these coupling schemes, the ideal cosine-cosine interaction exhibits the sparsest frequency landscape of multiphoton resonances.
However, when the frequency and nonlinearity of the auxiliary electromagnetic modes are not appropriately designed, a spectrum of new multiphoton transitions emerges.
Consequently, the benefit of nonlinear couplings is outweighed by the increasing number of leakage mechanisms in this enlarged Hilbert space. 
We experimentally validate these effects in a device implementing mediated cosine-cosine coupling and observe their impact on readout performance. 
We measure readout-induced leakage that varies by more than an order of magnitude over a $\lesssim7\%$ variation in readout frequency, demonstrating the intrinsic frequency sensitivity of this coupling scheme, similar to a linear coupling.

%Figure_1

Our results underscore that designing nonlinear-coupling architectures demands explicit modeling of all relevant electromagnetic modes and systematic frequency sweeps in both numerical simulations and spectroscopy experiments.
In this sense, engineering nonlinear couplings for qubit readout is analogous to high-order filter synthesis: additional modes provide greater control over the system response, but the performance becomes increasingly sensitive to mode frequencies, their nonlinearity, and inter-mode couplings.

%Figure_2
\begin{figure*}[t]
\includegraphics[width = \textwidth]{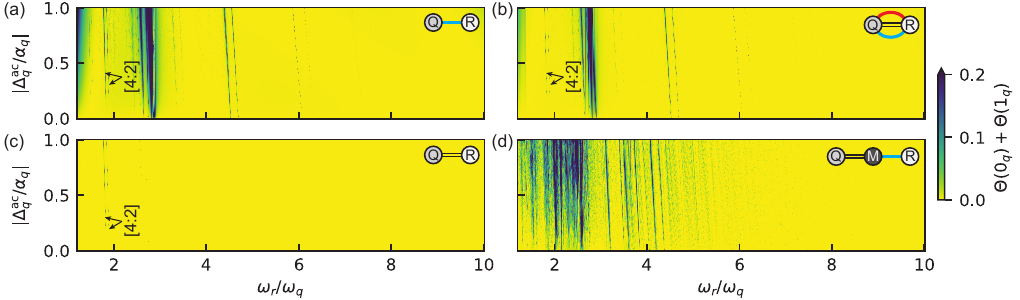}
\caption{Multiphoton resonance landscape visualized by the sum of displaced state overlaps $\Theta(0_t)+\Theta(1_t)$, calculated from both qubit states. The displaced-overlap is plotted against the readout drive frequency (normalized by qubit frequency) $\omega_d / \omega_q$ and readout power for different circuit realizations: (a) transmon linearly coupled to the drive~\cite{connolly_2025}, (b) balanced cross-Kerr interaction~\cite{chapple2025_bal} ($E_J\cos\hat\phi_q\cos\hat\phi_r + E_J\sin\hat\phi_q\sin\hat\phi_r- J\hat n_q\hat n_r$), (c) ideal cosine-cosine $\cos\hat\phi_q\cos\hat\phi_r$ coupling, and (d) effective nonlinear coupling mediated by an auxiliary mode (this work, see also Ref.~\cite{mori2025}). Although the ideal $\cos\hat\phi_q\cos\hat\phi_r$ interaction suppresses most multiphoton resonances, in  mediated $\cos\hat\phi_q\cos\hat\phi_r$ interaction, these resonances reappear. Their frequency and strength are governed by the mediator mode parameters, and in general, the mediated interaction increases frequency crowding.}
\label{fig:floquet_comparison} 
\end{figure*}

\textit{Multiphoton resonance in systems with nonlinear couplings:}\textemdash Designing the operator through which the drive couples to the effective qubit mode modifies the
\textit{allowed} nonlinear processes, their strengths, and resonance conditions.
For example, previous work~\cite{yao_aniket_2023} realized a parity-protected beamsplitter element by engineering the geometric symmetry of the Josephson device. In this case, half of the usual multiphoton processes were suppressed.
Similarly, several nonlinear coupling schemes have been engineered to realize effective dispersive interactions for qubit readout. 
We analyze three recently proposed qubit-readout interactions, and compare them to the standard case of linear coupling. These are: (i) ideal $\cos\hat\phi_q\cos\hat\phi_r$ coupling; (ii) balanced cross-Kerr interaction~\cite{chapple2025_bal}; (iii) effective  $\cos\hat\phi_q\cos\hat\phi_r$ coupling through an auxiliary mode (the subject of this work, see also Ref.~\cite{mori2025}). 

We perform Floquet steady-state simulations to quantitatively compare these cases. Across these simulations, the effective qubit frequency $\omega_q$ and anharmonicity $\alpha_q$ are held fixed to $\omega_{q}/2\pi=6.323$ GHz and $\alpha_{q}/2\pi=-182.2$ MHz. To reveal the frequency and power dependence of multiphoton resonances, the drive frequency is swept over a range $\omega_d/\omega_q \in \{1.2, 10\}$, and for each drive frequency, the drive amplitude is swept such that the induced ac Stark shift spans $|\Delta_{\rm ac}/\alpha_q|\in\{0, 1\}$. We plot the hybridization parameter $\Theta(j_t) = 1 - |\langle \tilde{j}_{t}(\xi,\omega_d)|\bar{j}_t(\xi,\omega_d)\rangle|^2$, which quantifies the deviation of a Floquet mode $\ket{\tilde{j}_t}$ from its ideal displaced state $\ket{\bar{j}_t}$ \cite{dai_hazra_weiss_2025, kurilovich2025}. In the absence of multiphoton resonances, $\Theta(j_t)$ remains near zero as the Floquet states only correspond to adiabatic displacements of the undriven states. However, when a multi-photon resonance is approached, $\Theta(j_t)$ increases, indicating a resonance condition. See Supplementary Materials~\cite{sm} for the simulated Hamiltonians and associated derivations.

As reported in previous works~\cite{chapple2024_long, chapple2025_bal, mori2025}, we observe that modifying the qubit-readout interaction Hamiltonian alters the strength of the multiphoton processes and even eliminates certain resonances. However, it does not guarantee that the qubit readout is free from leakage; instead, the impact depends on the choice of the drive frequency,
see Fig.~\ref{fig:floquet_comparison}. Remarkably, multi-photon transitions vanish in all cases for a high-frequency readout drive, due to the vanishing matrix elements of computational states with states near the top of the well \cite{kurilovich2025, connolly_2025}. However, how do the various coupling schemes compare at more conventional readout frequencies of $\omega_{r}/\omega_{q}\lesssim3$?

In ``Junction-readout'' through balanced cross-Kerr coupling \cite{chapple2025_bal, beaulieu2026_bal_exp}, the exchange (Jaynes–Cummings) coupling between transmon and readout is effectively eliminated by setting the $|0_t1_r\rangle \leftrightarrow |1_t0_r\rangle$ matrix element of the interaction Hamiltonian to zero. This modification of the interaction term also rescales the higher-order mixing terms and, when optimized, dilutes the strengths of nonlinear processes. This dilution can be observed by comparing the magnitude and width of lines in panels (a) and (b) of Fig.~\ref{fig:floquet_comparison}. Indeed, in the ideal case of perfect cancellation, the drive can no longer induce Rabi oscillations between $|0_{t}\rangle$ and $|1_{t}\rangle$ (note the relative absence of a background in panel (b) near $\omega_{r}/\omega_{q}\gtrsim 1$). But the symmetry of the Hamiltonian is otherwise unaffected: all other processes appearing in standard dispersive readout [panel (a)] are still present for junction readout. Thus, while balanced cross-Kerr coupling indeed suppresses the strength of multiphoton resonances, it does not eliminate them and remains vulnerable to the same leakage mechanisms as the standard linear coupling.

In contrast, the ideal $\cos\hat\phi_q\cos\hat\phi_r$ coupling significantly reduces the number of multiphoton resonances in the spectrum of unwanted processes owing to the imposed selection rule, see Fig.~\ref{fig:floquet_comparison}(c). However, even in this case the qubit is still susceptible to the \textit{allowed} $[m:n]$ processes, such that $\{m, n\}\in 2\mathbb N$ where $n$ drive quanta excite the qubit by $m$ levels. Especially, in this case, the so-called ``$[4:2]$'' resonance is a dominant multiphoton process; see the line on the left-hand side of Fig.~\ref{fig:floquet_comparison}(b), where two drive photons are converted into four transmon excitations. Thus, in engineering parametric gates or readout, one must avoid such lower-order ``allowed'' resonances, even for the most ideal coupling in circuits.

Further, in experiments, the $\cos\hat\phi_q\cos\hat\phi_r$ coupling is often achieved as an effective interaction mediated via an auxiliary mode. In this implementation, the two nonlinear modes of a multimodal circuit~\cite{gambetta_2011_dimon} share a $\cos\hat\phi_q\cos\hat\phi_m$ coupling, and one of these modes (auxiliary or mediator mode) is linearly hybridized with a readout mode, and the other nonlinear mode is treated as the qubit mode.
Due to the symmetry, only the mediator mode is driven; the drive cannot induce Rabi oscillations in the qubit mode. Even when this auxiliary mode is not intentionally engineered, additional modes appear in the circuit due to parasitic capacitances~\cite{salunkhe_2025}.
Further, the qubit often couples to several other intended and unintended circuit degrees of freedom on the chip~\cite{connolly_2025} or in the package~\cite{dai_hazra_weiss_2025, sheldon_2017_package_modes}.
We show that the presence of additional modes often spoils the benefit of the nonlinear coupling, and
a suboptimal mediated coupling exhibits the densest spectrum of nonlinear processes among the four qubit–resonator coupling schemes, as shown in Fig.~\ref{fig:floquet_comparison}(d). 
The multiphoton resonances in these implementations are sensitive to the frequency and anharmonicity of the auxiliary mode (here $\omega_{m}/2\pi=4.614$ GHz, $\alpha_{m}/2\pi=-110.4$ MHz, as realized in Ref~\cite{hazra_dai_2025}). Optimizing these parameters redistributes the resonances in frequency space and opens windows for readout frequencies free from unwanted transitions~\cite{Dimon_opt}. Such optimization must account for any unintended package modes that couple to either the qubit or the auxiliary mode.

%Figure_3
\begin{figure}[t!]
    \centering
    \includegraphics[width=\columnwidth]{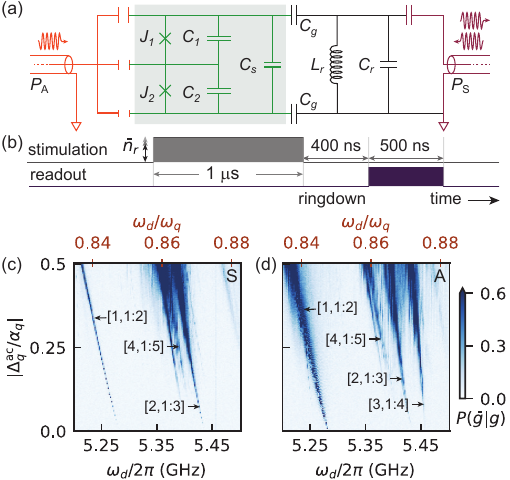}
    \caption{Impact of auxiliary modes and residual asymmetry. (a) Simplified lumped element model of the dimon-readout system and schematic comparison between the symmetric and asymmetric drives through ports $P_{\rm S}$ and $P_{\rm A}$, respectively. 
    (b)  Pulse sequence for characterizing drive-induced unwanted state transitions. After preparing the qubit in $\ket{g}$, a  $1 \mu$s stimulation tone of variable frequency and power is applied, followed by a readout to measure the escaped population, $P(\bar g|g)$.
    (c-d) Plotted $P(\bar g|g)$ as a function of drive frequency and drive power under symmetric and asymmetric drives on the device. (c) Auxiliary modes introduce multiple joint-excitation channels in the driven circuit as labeled by the shorthand notation $[x,y:n]$. (d) Under asymmetric drive, additional multiquanta resonances, such as $[3,1:4]$ and $[1,1:2]$, become prominent as the selection rule suppressing the forbidden transitions is now relaxed.}
    \label{fig:symmetry_dependence}
\end{figure}

\textit{Impact of auxiliary modes.}\textemdash We focus on an experimental multimodal device~\cite{roy_2017, pfeiffer_2024}, \textit{dimon}~\cite{hazra_dai_2025},  which implements a mediated effective cosine-cosine coupling. 
The lumped element model of the device is shown in Fig.~\ref{fig:symmetry_dependence}(a), showing the dimon circuit (green), readout resonator (black), symmetric drive line for readout (brown), and an asymmetric drive line for qubit control (orange). 
The qubit mode is the quadrupolar excitation of the multimode circuit, which is linearly uncoupled from the readout mode.
Elimination of the linear coupling between the qubit and the readout resonator results in intrinsic Purcell protection while maintaining dispersive interaction~\cite{hazra_dai_2025}.
The measured qubit and mediator mode frequencies are $6.231$ GHz and $4.605$ GHz, respectively~\cite{sm}.

Intuitively, for every auxiliary mode, the size of the Hilbert space expands, introducing more pathways for multiphoton transitions. Although the geometric symmetry imposes selection rules that forbid a subset of these processes, the overall effect is a more crowded frequency landscape, with an increasing number of multiphoton resonances. We demonstrate these transitions through a pump-probe spectroscopy of the circuit as shown in Fig.~\ref{fig:symmetry_dependence}(b). 
We first prepare the qubit in the ground state, and then apply a $1\,\mu{\rm s}$ long stimulation pulse with a varying power and frequency, and finally apply a readout pulse to measure escaped population from the initial state. To experimentally distinguish the processes that are ``forbidden'' from the ``allowed'' ones, we send the drive through both symmetric and asymmetric drive ports. In each case, the drive power is independently calibrated through the induced ac Stark shift at each drive frequency. In Fig.~\ref{fig:symmetry_dependence}(c) and (d), we plot the results for the two cases, respectively, across a frequency window displaying several multiphoton resonances in the circuit. The spectroscopy results, from both ground and excited states of the qubit, over a wider range of frequency, are presented in the Supplementary Materials~\cite{sm}.

To identify these transitions, we fit the experimentally measured parameters to the model circuit Hamiltonian and perform a Floquet steady-state simulation. 
From the branch analysis, we label the dominant transitions visible in this window with a notation $[x,y:n]$ denoting a transition $|0_q0_m\rangle \rightarrow |x_qy_m\rangle$ through the absorption of $n$ photons from the drive. Here, the state indices $q$ and $m$ denote the qubit and the auxiliary mode, respectively. All the transitions observed in this frequency window involve the auxiliary-mode states.
Thus, the auxiliary mode mediating the effective $\cos \hat\phi_q\cos\hat\phi_r$ interaction also introduces multiple pathways for multiphoton transitions. 
While symmetry suppresses processes like $[1,1:2]$ and $[3,1:4]$, allowed transitions, such as $[2,1:3]$ and $[4,1:5]$, lead to frequency crowding and could increase the leakage probability in the driven system. Notably, linearizing the auxiliary mode~\cite{mori2025} will suppress the matrix element for the higher-order transitions, thereby weakening their impact.

%Figure_4
\begin{figure}[t]
\includegraphics[width =\columnwidth]{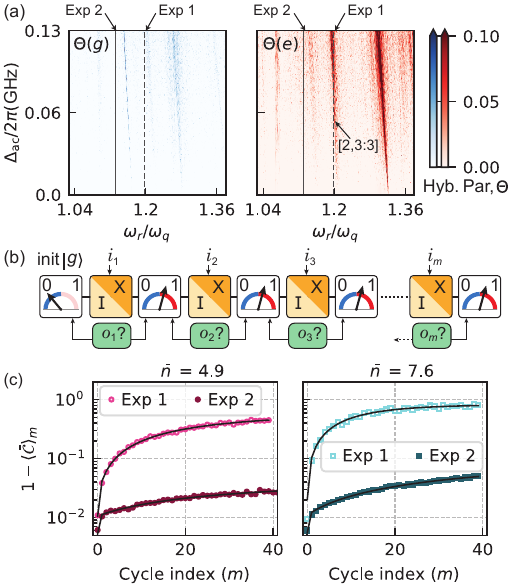}
\caption{
Readout-induced leakage in two experiments with different readout frequencies. (a) Simulated hybridization parameter $\Theta$,  from the qubit ground (blue) and excited (red) states, plotted against the ratio $\omega_r/\omega_q$ (x-axis) and the readout drive power (y-axis). In Exp 1 (vertical dashed line), a multiphoton resonance $[2,3:3]$ is activated from the excited state at the readout frequency; in Exp 2 (vertical solid line), this resonance is avoided by detuning the readout frequency. 
(b) Pulse sequence for benchmarking the readout-induced leakage. The readout operations are intertwined with a stochastic sequence of qubit operations consisting of identity ($\rm I$) and bit-flip ($\rm X$) operations. The measured binary readout record is then compared with the expected outcome. 
(c) Measured average error ($1-\langle \bar C\rangle_m$) in correlation between the stochastic input sequence and the inferred sequence from the readout results, plotted as a function of sequence length. Results are compared for two different readout powers,  $\bar{n}_r = \{4.9, 7.6\}$. Pale-hollow and dark-solid markers denote measurements from Exp 1 and Exp 2, respectively.
}
\label{fig:results_rilb} 
\end{figure}

\textit{Sensitivity to the choice of readout frequency.}\textemdash If these accidental resonance conditions are met, the multiphoton transitions introduced by the auxiliary mode cause readout-induced leakage. This situation is worsened if the qubit or the auxiliary mode couples to any other unintended modes, such as the chain modes in a Josephson junction array~\cite{singh_2025_jja}, package modes~\cite{sheldon_2017_package_modes,connolly_2025,dai_hazra_weiss_2025}, or modes arising from parasitic capacitances~\cite{salunkhe_2025}.

We illustrate this detrimental impact by performing a Floquet simulation, as shown in Fig.~\ref{fig:results_rilb}(a), around $\omega_r/2\pi = 7.5$ GHz, in a range where we aim to place the readout resonator. As before, we scale the drive frequency axis by the qubit frequency. We find that in our device, around $\omega_r/\omega_q \approx 1.2$, the readout is impacted by an $8$-wave mixing process exciting $|1_q0_m\rangle\leftrightarrow|3_q3_m\rangle$ by absorbing $3$ drive photons. Similar to the transitions in Fig.~\ref{fig:symmetry_dependence}, this multiphoton resonance exists only due to the mediator mode. Importantly, the Floquet spectrum in Fig.~\ref{fig:results_rilb}(a) shows that the onset of this multiphoton resonance is highly sensitive to the choice of drive frequency. As a result, the drive power required to trigger leakage varies by up to an order of magnitude across the explored frequency window, even though the system parameters and coupling strengths remain unchanged. Characterizing leakage at a single readout frequency is therefore insufficient to assess a coupling scheme. 

To demonstrate the sensitivity of readout-induced leakage to the readout frequency, we perform two independent readout experiments on the same device in successive cooldowns. The readout resonator frequencies are $7.513$ and $7.025$ GHz, respectively, differing by less than $7\%$. The dispersive shift $\chi$ and the resonator linewidth $\kappa$ are kept nominally similar to ensure a comparable readout signal-to-noise ratio (SNR) at equal readout photon number $\bar n_r$. These two configurations are labeled `\texttt{Exp\,1}' and `\texttt{Exp\,2}', and the corresponding readout frequencies are shown respectively by the vertical solid and dashed lines in Fig.~\ref{fig:results_rilb}(a).
For each of these two configurations, we calibrate the readout pulse including an active resonator emptying sequence~\cite{mcclure2016_clear,hazra_dai_2025, chatterjee2025_rl} for two different readout powers, $\bar n_r=4.9$ and $\bar n_r=7.6$. We independently optimize the duration, detuning, and shape of the pulse to maximize fidelity for each readout setting.

\setlength{\fboxsep}{1pt}
For these two readout frequencies, we characterize the readout-induced leakage for repeated readout operations~\cite{hazra_dai_2025} at the two different readout powers. The pulse sequence of this experiment is shown in Fig.~\ref{fig:results_rilb}(b). It consists of an interleaved sequence of qubit control, $\rm X$ or $\rm I$,  and readout operations.
We measure the correlation between the applied qubit operations and the readout outcomes, and 
from the measured decay of correlation, we simultaneously extract the leakage ($L_\uparrow$) and seepage ($L_\downarrow$) probability per readout, and the readout fidelity, $\mathcal{F}_{ge}$, assuming an incoherent leakage model~\cite{sm}:
\begin{equation}
\begin{split}
    \langle\bar{\mathcal{C}}_{m=0}\rangle 
    &= \left(2\mathcal{F}_{ge}-1\right) \quad\\
    \langle\bar{\mathcal{C}}_{m>0}\rangle&= \frac{L_\downarrow+L_\uparrow(1-L_\downarrow-L_\uparrow)^m}{L_\downarrow+L_\uparrow}\left(2\mathcal{F}_{ge}-1\right)^2
\end{split}
\label{eq:rilb_trajectory_main}
\end{equation}
Fig.~\ref{fig:results_rilb}(c) shows the measured correlation as a function of the number of readout pulses. 
The readout parameters and extracted leakage and seepage rates are reported in Table~\ref{tab:rilb_before_after}, with the leakage rates showing more than an order of magnitude difference between the two experiments.

%Table_1
\begin{table}[t]
    \centering
    \begin{tabular}{ m{3cm} m{1.2cm} m{1.2cm} m{1.2cm} m{1.2cm}}
        \toprule
        Readout power &  \multicolumn{2}{c}{$\bar{n}_r = 4.9$} & \multicolumn{2}{c}{$\bar{n}_r = 7.6$}\\
        \midrule
        {}  & Exp 1 & Exp 2 &  Exp 1 & Exp 2 \\
        Duration (ns)   & $120$ & $104$ & $100$ & $100$ \\
        Fidelity (\%)  & $99.5$ & $99.7$ & $99.0$ & $99.7$ \\
        Leakage (\%)  & $2.1$ & $0.1$ & $6.3$ & $0.1$ \\
        Seepage (\%)  & $1.6$ & $2.3$  & $1.4$ & $1.3$ \\
        \bottomrule
    \end{tabular}
    \caption{Quantitative comparison of readout fidelity, readout-induced leakage and seepage rates, before (Exp 1) and after (Exp 2) the optimization of readout frequency, each characterized for two different readout powers.}
    \label{tab:rilb_before_after}
\end{table}

\emph{Conclusions.}\textemdash Our results highlight that, in principle, qubit readout based on native nonlinear coupling, especially cosine-cosine coupling, improves the quantum non-demolition character of the readout for a larger span of frequencies compared to standard linear coupling. In practice, however, auxiliary degrees of freedom, which are often essential to achieve this improved interaction, introduce additional leakage channels. Since such modes are often unavoidable in realistic implementations, they must be explicitly included when modeling and optimizing the device. 
Finally, since multiphoton resonances are particularly sensitive to drive frequency, 
readout performance at a single frequency is insufficient to evaluate nonlinear coupling schemes.
A comprehensive assessment, therefore, requires frequency sweeps in both experimental and theoretical characterization.

\begin{acknowledgments}
The authors acknowledge Thomas Connolly,  Andy Z. Ding, Vidul R. Joshi, Akshay Koottandavida, Pavel D. Kurilovich, Aniket Maiti, Alessandro Miano,  Jaya Venkatraman, Xu Xiao, and Yao Lu for insightful discussions. 
S.H. thanks Alexandru Petrescu for stimulating discussions. 
This research was sponsored by the Army Research Office (ARO) under grant nos. W911NF-23-1-0051, by the Air Force Office of Scientific Research (AFOSR) under grants FA9550-19-1-0399 and FA9550-21-1-0209, and by the U.S. Department of Energy (DoE), Office of Science, National Quantum Information Science Research Centers, Co-design Center for Quantum Advantage (C2QA) under contract number DE-SC0012704. The views and conclusions contained in this document are those of the authors and should not be interpreted as representing the official policies, either expressed or implied, of the ARO, AFOSR, DoE or the US Government. The US Government is authorized to reproduce and distribute reprints for Government purposes, notwithstanding any copyright notation herein. Fabrication facilities use was supported by the Yale Institute for Nanoscience and Quantum Engineering (YINQE) and the Yale SEAS Cleanroom.
L.F. is a consultant and shareholder of D-Wave Quantum.
D.~K.~W thanks the Yale Center for Research Computing, specifically Thomas Langford and Aya Nawano, for guidance and assistance in computations run on the Grace and Bouchet clusters.
\end{acknowledgments}

\textit{Author contribution}\textemdash S.H. and W.D. designed the experiment and built the measurement setup.
S.H. characterized the device, performed the multiphoton resonance (DUST) spectroscopy, and analyzed all the data. 
W.D performed the leakage benchmarking experiment. 
D.K.W. performed all the numerical simulations. 
P.D.P. contributed to state labeling and transition identification.
L.F. and M.H.D. administered and supervised the project.
S.H., D.K.W, and W.D. wrote the manuscript with inputs from all coauthors.

%\noindent The data that support the findings of this letter are openly available.

% \bibliographystyle{apsrev4-2}
% \bibliography{reference}
%%%%%%%%%%%%%%%%%%%%%%%%%%%%%%%%%%%%%%%%%%%%%%%%%%%%%%%
% bbl file

%%%%%%%%%%%%%%%%%%%%%%%%%%%%%%%%%%%%%%%%%%%%%%%%%%%%%%
%%% Supplementary materials

\onecolumngrid
\newpage

%%%%%%%%%%
\setcounter{section}{0}
\setcounter{table}{0}
\setcounter{figure}{0}
\setcounter{page}{1}
\renewcommand{\thepage}{S-\arabic{page}}
\thispagestyle{empty}
\renewcommand{\thefigure}{S\arabic{figure}}
\renewcommand{\thetable}{S\arabic{table}}
\renewcommand{\theequation}{S\arabic{equation}}
\renewcommand{\thesection}{S\arabic{section}}
\renewcommand{\thesubsection}{S\arabic{section}.\arabic{subsection}}

\begin{center}
{\Large\bfseries Supplementary Materials for ``Readout-Induced Leakage in Superconducting Circuits with Nonlinear Couplings''}
\end{center}

% \tableofcontents

\section{Model Hamiltonian for nonlinear couplings}
In this section, we define the model Hamiltonian for each of the different nonlinear coupling schemes described in the main text. We also explain the labeling scheme for the multimode system and how we identify the transitions.
\subsection{Native cosine-cosine coupling}

Here we consider the most ideal scheme, where the readout resonator is coupled to the qubit only via a cosine-cosine interaction. This is effectively the case of ``longitudinal readout", see Refs.~\cite{_didier_2015, _chapple2024_long}. The Hamiltonian is
\begin{align}
H_{cc} &= 4E_{C}(\nt - n_{g})^2 - E_{J}\cos(\phit)\cos(\phir) + \omega_{r}\adagop\aop +A\nr\cos(\omega_{d}t), \\ \nonumber 
\end{align}
where
$\phir = \phi_{r}(\aop + \adagop)$, $\nr = -in_{r}(\aop - \adagop)$, $\phi_{r}=(8E_{Cr}/E_{L})^{1/4}/\sqrt{2}$, $n_{r}=(8E_{Cr}/E_{L})^{-1/4}/\sqrt{2}$ and $E_{Cr}, E_{L}$ are the self capacitance and inductance energies of the readout resonator.
To eliminate the linear drive, we utilize the unitary transformation $\hat U=\exp(\alpha\adagop - \alpha^{*}\aop)$, with
\begin{align}
\alpha = \frac{An_{r}(-i\omega_{r}\cos(\omega_{d}t)-\omega_{d}\sin(\omega_{d}t))}{\omega_{r}^2-\omega_{d}^2}.
\end{align} 
The transformed Hamiltonian is
\begin{align}
\label{eq:H_cc}
H'_{cc} &=U^{\dagger}H U-iU^{\dagger}\dot U \\ \nonumber &= 4E_{C}(\hat n - n_{g})^2 - E_{J}\cos(\hat \phi)\cos(\xi\sin(\omega_{d}t)),
\end{align}
c.~f.~ Eq.(J1) of Ref.~\cite{_mori2025}, where $\xi=\tfrac{A\omega_{d}}{\omega_{r}^2-\omega_{d}^2}$ and we have neglected quantum fluctuations of the readout resonator. This is the Hamiltonian used for Floquet analysis of ideal cosine-cosine coupling.

\subsection{Mediated cosine-cosine coupling}

It is possible to engineer an effective cosine-cosine coupling by introducing an auxiliary mediator mode \cite{_hazra_dai_2025, _mori2025}. Here we consider the driven dimon+resonator system introduced in Ref.~\cite{_hazra_dai_2025}. The Hamiltonian is \cite{_hazra_dai_2025}
\begin{align}
H_{D} &= 4E_{Cq}(\hat n_{q}-n_{gq})^2 + 4E_{Cm}(\hat n_{m}-n_{gm})^2 -2E_{J}\cos(\hat\phi_{m})\cos(\hat\phi_{q}) \\ \nonumber &\quad+\omega_{r}\adagop\aop + g\hat n_{m}\nr+A\nr\cos(\omega_{d}t),
\end{align}
where, due to the symmetry of the device, only the dipolar mediator mode linearly couples to the readout resonator. 
As above, we move into a displaced frame and neglect quantum fluctuations of the resonator, yielding
\begin{align}
\label{eq:H_dimon}
H_{D}' = 4E_{Cq}(\hat n_{q}-n_{gq})^2 + 4E_{Cm}(\hat n_{m}-n_{gm})^2 -2E_{J}\cos(\hat\phi_{m})\cos(\hat\phi_{q}) - B \hat n_{m} \cos(\omega_{d}t),
\end{align}
where $B=\tfrac{2gn_{r}^2A\omega_{r}}{\omega_{r}^2-\omega_{d}^2}$. This is the Hamiltonian used for Floquet analysis of effective cosine-cosine readout via a mediator mode. In this work, we take $E_{J}=16.52$ GHz, $E_{CJ}= 0.3221$ GHz, and $E_{Cs}=0.775$ GHz. Noting the effective charging energy for the two modes are given by $E_{Cq}=\tfrac{E_{CJ}}{2}$ and $E_{Cm}=\tfrac{E_{CJ}E_{Cs}}{2(E_{Cs} + 2E_{CJ})}$ \cite{_hazra_dai_2025}, we obtain $E_{Cq} = 0.161$ GHz, $E_{Cm}=0.088$ GHz. 
We note that Eq.~\eqref{eq:H_cc} can be approximately obtained from Eq.~\eqref{eq:H_dimon} by performing another displacement transformation on the mediator mode and neglecting the nonlinearity and quantum fluctuations associated with that degree of freedom \cite{_mori2025}. 

Only a specific class of processes is allowable for the driven dimon system. Because the mediator mode is directly displaced, it is only the sum of the number of drive photons plus the change in the number of mediator excitations that must be even. That is, the excitation number of the mediator mode can change by an odd number if an odd number of drive photons are involved. 
Conversely, in every allowed nonlinear resonance, the change in the number of qubit mode excitations must be even.

\subsection{Balanced cross-Kerr coupling (Junction readout)}
In this scheme, the transmon is coupled to the readout resonator via a Josephson junction in parallel to the standard coupling capacitor \cite{_chapple2025_bal}. Here, we derive an analytic expression for the cancellation condition as well as the single-mode Hamiltonian used for the Floquet analysis. The Hamiltonian is [c.~f. Eq.(1) of Ref.~\cite{_chapple2025_bal}]
\begin{align}
H &= 4E_{C}(\nt - n_{g})^2 - E_{J}\cos\phit + \omega_{r}\adagop\aop - E_{Jc}\cos(\phit - \phir)+J\nt\nr -iA'(\aop-\adagop)\cos(\omega_{d}t),
\end{align}
where, for ease of later notation, we have defined $A'=n_{r}A$.
Again, moving into a displaced frame and neglecting quantum fluctuations of the cavity, we obtain

\begin{align}
H' 
\label{eq:H_jr}
&=4E_{C}(\nt - n_{g}^2) - E_{J\Sigma}\cos\phit 
\\\nonumber&\quad 
- E_{Jc}\cos\phit\{\cos(\phi_{r}\xi\sin[\omega_{d}t])-1\}+ E_{Jc}\sin\phit\sin(\phi_{r}\xi\sin[\omega_{d}t])-J\nt n_{r}\zeta\cos(\omega_{d}t),
\end{align}
where 
\begin{align}
\xi=\frac{2A'\omega_{d}}{\omega_{r}^2-\omega_{d}^2}, \quad\zeta=\frac{2A'\omega_{r}}{\omega_{r}^2-\omega_{d}^2}, \quad E_{J\Sigma}=E_{J}+E_{Jc},
\end{align}
and we have extracted out the time-independent part of the Hamiltonian. This is the Hamiltonian used for Floquet analysis of junction readout, where we take $\xi=\zeta$ under the assumption that the resonator is nearly resonantly driven.

\subsection{Identifying transitions}

We identify the levels involved in Dimon transitions based on a normal-mode analysis. Expressions for the normal modes of the Dimon circuit in the symmetric case $E_{J1}=E_{J2}=E_{J}$ and $C_{1}=C_{2}=C$ were previously given in Ref.~\cite{_hazra_dai_2025}. Here, we utilize a robust formalism for identifying the circuit normal modes even in the presence of parameter asymmetries. We develop the formalism in full generality, following closely the analysis presented in Ref.~\cite{_Weiss2021}. We then apply it to the case of dimon. We consider a generic circuit-QED Lagrangian
\begin{align}
\label{eq:Lagrangian}
\mathcal{L}=\frac{1}{2}\dot{\vec{\Phi}}^{T}\mathcal{C}\dot{\vec{\Phi}} -U(\vec{\Phi}; \vec{\Phi}_{\rm ext}),
\end{align}
where $\vec{\Phi}$ is a vector of the node fluxes of the circuit, $\mathcal{C}$ is the capacitance matrix, and $\vec{\Phi}_{\rm ext}$ is the vector of external fluxes threaded through various loops. We first identify a local minimum of the potential to expand around, where the quadratic terms of the Lagrangian expanded around this point will define our normal modes. We take this point to be the origin without loss of generality. The Lagrangian expanded up to second order is
\begin{align}
\label{eq:second_order_lagrangian}
\mathcal{L}^{(2)}=\frac{1}{2}\dot{\vec{\Phi}}^{T}\mathcal{C}\dot{\vec{\Phi}} - \frac{1}{2}\vec{\Phi}^{T}\Gamma\vec{\Phi},
\end{align}
where we have defined the inverse inductance matrix
\begin{align}
\Gamma_{ij} = \partial_{\Phi_{i}}\partial_{\Phi_{j}}U|_{\vec{0}}.
\end{align}
We now obtain the normal modes of Eq.~\eqref{eq:second_order_lagrangian} using elementary techniques \cite{_Goldstein}. We assume an oscillatory ansatz $\vec{\Phi} = \vec{\xi}_{\mu}e^{-i\omega_{\mu}t}$, where $\mu$ indexes the normal modes (and there is no summation on $\mu$). Inserting this definition into the equation of motion for Eq.~\eqref{eq:second_order_lagrangian} yields the generalized eigenvalue problem
\begin{align}
\label{eq:generalized_eigenvalue}
\Gamma\vec{\xi}_{\mu}=\omega_{\mu}^2C\vec{\xi}_{\mu}.
\end{align}
We may now solve for the normal-mode eigenvectors $\vec{\xi}_{\mu}$ and normal-mode frequencies $\omega_{\mu}$. Collecting all of the eigenmode vectors into a matrix $\Xi=(\vec{\xi}_{\mu})$, the normal-mode variables $\vec{\Theta}$ are related to the original variables $\vec{\Phi}$ by the transformation $\vec{\Phi}=\Xi\vec{\Theta}$. The second-order Lagrangian is now diagonal in the $\Theta_{\mu}$ variables
\begin{align}
\mathcal{L}^{(2)} = \frac{1}{2}\sum_{\mu}[c_{\mu}\dot\Theta_{\mu}^2-\omega_{\mu}^2c_{\mu}\Theta_{\mu}^2],
\end{align}
where $\vec{\xi}_{\mu}^{\, T}\mathcal{C}\vec{\xi}_{\mu}=c_{\mu}$ defines the normalization of the normal-mode eigenvectors. 
Legendre transform to the second-order Hamiltonian yields
\begin{align}
H^{(2)} &= \frac{1}{2}\sum_{\mu}\left[\frac{\mathcal{Q}^2}{c_{\mu}} + \omega_{\mu}^2c_{\mu}\Theta_{\mu}^2 \right] \\ \nonumber 
&= \frac{1}{2}\sum_{\mu}\left[ \frac{(2e)^2}{c_{\mu}}p_{\mu}^2 + \omega_{\mu}^2c_{\mu}\left(\frac{\Phi_{0}}{2\pi}\right)^2\theta_{\mu}^2\right] \\ \nonumber 
&= \frac{1}{2}\sum_{\mu}\hbar\omega_{\mu}(p_{\mu}^2+\theta_{\mu}^2),
\end{align}
where we have introduced the conjugate momentum $\mathcal{Q}=2ep_{\mu}=\frac{\partial L^{(2)}}{\partial \Theta_{\mu}}$ as well as the reduced normal-mode flux $\vec{\theta}=\frac{2\pi}{\Phi_{0}}\vec{\Theta}$. In the third line, we have made the convenient choice $c_{\mu}=(2e)^2/(\hbar\omega_{\mu})$ \cite{_Weiss2021}. 
This quadratic Hamiltonian is now immediately diagonalized by introducing creation and annihilation operators, e.g. $\theta_{\mu}=\frac{1}{\sqrt{2}}(a_{\mu} + a_{\mu}^{\dagger})$, with $[\hat{a}_{\mu}, \hat{a}_{\nu}^{\dagger}]=\delta_{\mu \nu}$.

We may now use the normal-mode expansion in the (unexpanded) Hamiltonian
\begin{align}
\hat{\mathcal{H}}=4\hat{\vec{n}}^{T}E_{C}\hat{\vec{n}}+U(\hat{\vec{\phi}}; \Phi_{\rm ext}),
\end{align}
obtained by taking the Legendre transform and quantizing from Eq.~\eqref{eq:Lagrangian}. We have defined $\vec{Q}=2e\vec{n}=\partial \mathcal{L}/\partial\dot{\vec{\Phi}}$ as well as $\vec{\phi}=\frac{2\pi}{\Phi_{0}}\vec{\Phi}$. The original variables can now be written in terms of the normal-mode variables as
\begin{align}
\hat{\vec{\phi}} = \Xi\hat{\vec{\theta}}=\frac{1}{\sqrt{2}}\Xi(\hat{\vec{a}}+\hat{\vec{a}}^{\dagger}), \\ \nonumber
\hat{\vec{n}}=\frac{-i}{\sqrt{2}}\Xi^{-T}(\hat{\vec{a}}-\hat{\vec{a}}^{\dagger}),
\end{align}
where $\Xi^{-T}=(\Xi^{-1})^{T}$.

\vspace{1em}

\noindent \textbf{Application to transmon:}
In the case of the Transmon, the capacitance and inverse-inductance matrices are scalars given by $C,(2\pi/\Phi_{0})^2E_{J}$, respectively. We then readily obtain $\Xi=(8E_{C}/E_{J})^{1/4}$, in agreement with the known expression for the zero-point fluctuations \cite{_Koch2007, _Minev2021}.

\vspace{1em}

\noindent \textbf{Application to dimon:}
The Lagrangian of the Dimon is \cite{_hazra_dai_2025}
\begin{align}
\mathcal{L} &=\frac{1}{2}\left[C_{1}\dot{\Phi}_{1}^2+C_{2}\dot{\Phi}_{2}^2+C_{s}(\dot{\Phi}_{2}-\dot{\Phi}_{1})^2 \right] \\ \nonumber &\quad + E_{J1}\cos(2\pi\frac{\Phi_{1}}{\Phi_{0}})+E_{J2}\cos(2\pi\frac{\Phi_{2}}{\Phi_{0}}).
\end{align}
The capacitance and inverse inductance matrices are
\begin{align}
C &=\left(
\begin{matrix}
C_{1}+C_{s} & -C_{s} \\ -C_{s} & C_{2} + C_{s}
\end{matrix}
\right) \\ 
\Gamma &= \left(\frac{2\pi}{\Phi_{0}}\right)^2\left(\begin{matrix}
    E_{J1} & 0 \\ 0 & E_{J2}
\end{matrix} \right).
\end{align}
These matrices define the normal-mode matrix $\Xi$ via Eq.~\eqref{eq:generalized_eigenvalue}, and we obtain
\begin{align}
\label{eq:harmonic_hamiltonian}
H &= \sum_{\mu=q,m}\hbar\omega_{\mu}a_{\mu}^{\dagger}a_{\mu} - \sum_{j=1,2}E_{Jj}\left[\cos(\phi_{j})+\phi_{j}^2/2 \right],
\end{align}
where $\phi_{j}=\frac{1}{\sqrt{2}}\sum_{\mu}\Xi_{j\mu}(a_{\mu}+a_{\mu}^{\dagger})$. In the case of symmetric parameters $C_{1}=C_{2}=C_{J}$, $E_{J1}=E_{J2}=E_{J}$, we obtain $\omega_{q}=\sqrt{16E_{Cq}E_{J}}, \omega_{m}=\sqrt{16E_{Cm}E_{J}}$, where $E_{Cq}=\frac{e^2}{2(2C_{J})}$ and $E_{Cm}=\frac{e^2(E_{CJ}E_{Cs})}{2(2E_{CJ}+E_{Cs})}$ in agreement with previously derived results \cite{_hazra_dai_2025}. The normal-mode matrix is
\begin{align}
\Xi = \left(\begin{matrix}
    \xi_{m} & \xi_{q} \\ -\xi_{m} & \xi_{q}
\end{matrix} \right),
\end{align}
where $\xi_{m}=\left(\frac{4E_{Cm}}{E_{J}}\right)^{1/4}$ and $\xi_{q}=\left(\frac{4E_{Cq}}{E_{J}}\right)^{1/4}$. The Hamiltonian can now be written in the normal-mode basis as
\begin{align}
\label{eq:H_normal_mode}
H &= \sum_{\mu=q,m}(\hbar\omega_{\mu}\hat{a}_{\mu}^{\dagger}\hat{a}_{\mu} -E_{J}\xi_{\mu}^2\theta_{\mu}^2 ) \\ \nonumber &\quad- 2E_{J}\cos(\xi_{q}\hat{\theta}_{q})\cos(\xi_{m}\hat{\theta}_{m}).
\end{align}
Expanding it up to fourth order, we obtain
\begin{align}
\label{eq:H_4}
H^{(4)} &= \sum_{\mu=q,m}\left[\hbar\omega_{\mu}'\hat{a}_{\mu}^{\dagger}\hat{a}_{\mu} - \frac{E_{C\mu}}{2}(\hat{a}_{\mu}^{\dagger})^2\hat{a}_{\mu}^2\right] \\ \nonumber &\quad-2\sqrt{E_{Cm}E_{Cq}}\hat{a}_{m}^{\dagger}\hat{a}_{m}\hat{a}_{q}^{\dagger}\hat{a}_{q},
\end{align}
where $\omega_{\mu}'=\omega_{\mu}-\sqrt{E_{Cm}E_{Cq}}-E_{C\mu}$. We emphasize here that the usefulness of this formalism lies in its generality. In the presence of possible parameter asymmetry, analytical expressions become tedious; nevertheless, we may solve easily for the normal-mode frequencies and $\Xi$ matrix numerically.

We are typically interested in identifying the bare labels associated with high-lying states. Thus, in practice, we do not use the fourth-order Hamiltonian Eq.~\eqref{eq:H_4}, but rather diagonalize the more general one, Eq.~\eqref{eq:H_normal_mode}. 
Having obtained the eigenenergies and dressed eigenvectors, 
we compute the overlaps of the eigenvectors with the bare-product states $|j_{q},k_{m}\rangle$ (labeled by the number of harmonic excitations in each mode). We assign each eigenvector the label associated with the bare state with the largest overlap in magnitude. 

We can now identify which states (in terms of their bare labels) participate in a nonlinear transition by performing branch analysis and mapping the dressed indices of the involved states to their respective bare labels. We can then extract the number of drive photons by dividing the energy difference by the drive frequency (it is best to perform this analysis for small values of the drive amplitude, to avoid frequency differences changing due to Stark shifts).

\begin{figure}[t]
    \centering
    \includegraphics[width=0.45\textwidth]{ 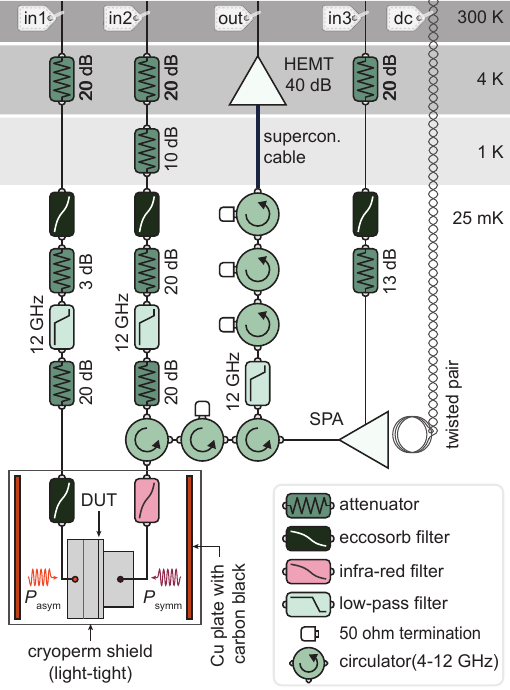}
    \caption{Cryogenic microwave setup for the experiments presented in the main text. The readout (in2) is carried out through the symmetrically coupled port, $P_{\rm symm}$, and the qubit drive (in1) is applied through the asymmetrically coupled port, $P_{\rm asym}$. The SPA is operated to produce nominally 20 dB gain, and the SPA pump is delivered through an additional input port (in3). All the input lines are equipped with Ecosorb filters to block high-frequency IR radiation. Within the light-tight shield, we put an additional Ecosorb filter for the driveline and a low-loss HERD\cite{_herd} IR filter for the readout line.}
    \label{fig:cryo_wiring}
\end{figure}

\section{Experimental setup and device parameters}
\label{appendix:device_params}
The experimental setup is identical to that of the previous work~\cite{_hazra_dai_2025}. The device is fabricated on a Heat Exchanger Method (HEM) sapphire chip with standard aluminum thin film fabrication. The junctions are formed through a bridge-free technique. The chip is placed symmetrically at the center of a 3D waveguide cavity made of 6061 aluminum alloy. The TE101 mode of the 3D waveguide cavity plays the role of the readout resonator. 
The symmetric drive is delivered through a waveguide adapter coupled to the readout cavity through an aperture. 
The asymmetric drive port is delivered through a drive pin placed below the chip, evanescently coupled to the capacitor pads of the device. The details of the assembly are illustrated in Fig.~\ref{fig:system_and_setup}(a), and an optical image of the fabricated device is shown in Fig.~\ref{fig:system_and_setup}(b). The device is then placed inside a light-tight Cryoperm shield protecting it from infrared radiation and stray magnetic fields. The details of the cryogenic filtering and attenuation are illustrated in Fig.~\ref{fig:cryo_wiring}. A SNAIL parametric amplifier (SPA) is installed in the output line to achieve near quantum-limited efficiency in the signal amplification chain. We have measured the temperature of the qubit mode $T_{q}= 51~{\rm mK}$ based on its thermal populations.

\subsection{Device properties}
The data shown in the main text consists of experiments from three different cooldowns, viz. the readout-induced leakage benchmarking before the readout frequency optimization is implemented in cooldown I, the characterization of the drive-induced unwanted transitions was performed in cooldown  II, and finally, the readout-induced leakage benchmarking experiment with the optimized readout is from cooldown III. 
The measured qubit frequencies are $\omega_q/2\pi = \{6.271, 6.231, 6.209\}$ GHz across the three cooldowns. Such frequency shifts are primarily attributed to the aging of the Josephson junction between successive cooldowns. The readout frequencies during the first two cooldowns are unchanged and are deliberately adjusted during the third cooldown. 
These frequencies are
$\omega_r/2\pi = \{7.513, 7.513, 7.025\}$ GHz. The measured cavity linewidth, and the dispersive shift of the readout cavity due to the qubit mode are $\kappa/2\pi = \{11.6, 11.6, 15.5\}$ MHz and $\chi/2\pi = \{-6.4,-6.4, -6.0\}$ MHz, respectively. Note that we intend to keep the cavity linewidth and the dispersive shift nominally unchanged while replacing the readout resonator between cooldown II and III. 

The relaxation time, Ramsey decoherence time, and the Hahn echo decoherence time of the qubit mode during the final cooldown (cooldown III) are measured to be $T_1 \sim 100\pm 10~\mu{\rm s}$, $T_2^R \sim  45\pm5~\mu{\rm s}$, and $T^E_2 \sim 90\pm 10~\mu{\rm s}$,  respectively, with the error bar referring to fluctuations of the measured values over days. Note that the qubit mode is intrinsically Purcell-protected by symmetry.
%2024-09-12

\section{Readout settings and pulse calibration} 
\begin{figure}[t]
    \centering
    \includegraphics[width=0.95\textwidth]{ 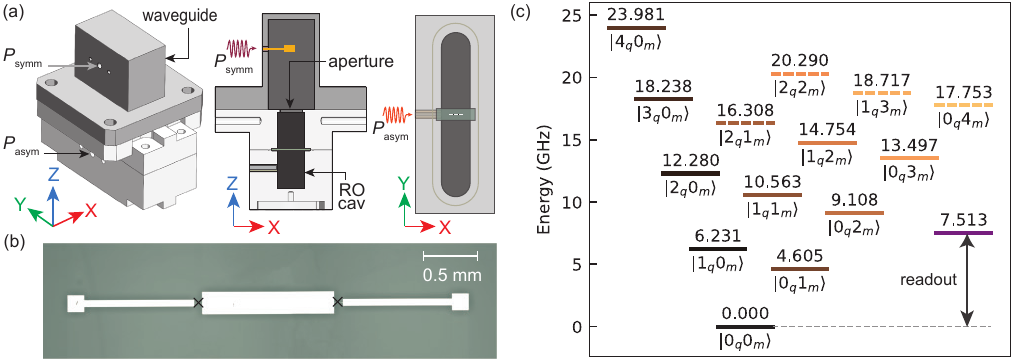}
    \caption{Experimental device and energy spectrum. (a) Schematic of the 3D cavity and the waveguide coupler for symmetric drive and an additional drive port for asymmetric drive.
    (b) A micrograph of the fabricated multimode device, consisting of two Josephson junctions (schematically represented by the black cross symbols).
    (c) Measured static (undriven) energy level diagram ($E_i$) of the qubit-mediator joint Hilbert space during cooldown II. The states are labeled as $|a_qb_m\rangle$, where $a$ and $b$ represent the number of excitations in the qubit and the mediator mode, respectively. The solid levels are characterized through the Ramsey (and two-tone spectroscopy) experiment; the dashed levels are characterized only via two-tone spectroscopy. The readout resonator frequency ($\omega_r/2\pi = 7.513$ GHz) is also shown in the diagram for reference.
    }
    \label{fig:system_and_setup}
\end{figure}

For device characterization and spectroscopy of drive-induced unwanted transitions, we perform a $500~\rm ns$ square-pulse readout in conjunction with phase-preserving amplification and multi-state discrimination. The energy levels of the qubit-mediator system, illustrated in Fig.~\ref{fig:system_and_setup}(c), are characterized through sequential state preparation, followed by two-tone spectroscopy and/or Ramsey interferometry. For each transition, the readout detuning is independently optimized to maximize the readout signal-to-noise ratio (SNR) between the levels involved in that transition.

For the high-fidelity repeated readout operations in the leakage benchmarking experiment, we employ a phase-sensitive amplification to maximize the efficiency of the amplification chain. The readout-induced leakage benchmarking experiment is performed during cooldown I and cooldown III.

\begin{table}[b]
    \centering
    \begin{tabular}{ m{3.6cm} m{3.4cm} m{4cm}}
        \toprule
        Experiment &  Durations (ns) & Relative amps \\
        \midrule
        $\bar n_r=4.9$ (Exp 1)   &  $\left[12,80,16,12\right]$ & $\left[2.81,  1.  , -3.35,  2.05\right]$ \\
        $\bar n_r=4.9$ (Exp 2)   &  $\left[8,76,12,8\right]$ & $\left[3.36,  1.  , -2.22,  0.92\right]$ \\
        \\
        $\bar n_r=7.6$ (Exp 1)   &  $\left[12,60,16,12\right]$ & $\left[2.82,  1.  , -3.36,  2.05\right]$ \\
        $\bar n_r=7.6$ (Exp 2)   &  $\left[8,72,12,8\right]$ & $\left[3.36,  1.  , -2.22,  0.92\right]$ \\
        \bottomrule
    \end{tabular}
    \caption{Parameters for the shaped readout pulse optimized separately for different powers in each of the two cooldowns.}
    \label{tab:ro_pulse_parameter}
\end{table}

\subsection{Calibration of active resonator reset pulse}
The readout-induced leakage benchmarking experiment characterizes leakage and seepage per readout by repeatedly applying the readout operation, interleaved with a stochastic sequence of identity and bit-flip operations on the qubit. Therefore, to eliminate the \textit{dead-time} after every readout operation (in which the photons in the resonator naturally decay into the bath), we apply an active resonator reset segment after each readout. The readout pulse contains four segments $\{A_i, \tau_i\},~i=1,\cdots,4$, with amplitudes $A_i$ and lengths $\tau_i$. The pulse begins with a sharp impulse, $\{A_1, \tau_1\}$, to overcome the transient behavior of the resonator, followed by a relatively longer, lower-amplitude segment $\{A_2, \tau_2\}$, driving the resonator towards a steady state, and then a two-segment active reset drive, $\{A_3, \tau_3\}$, and $\{A_4, \tau_4\}$. The lengths, $\tau_i$, and relative amplitudes, $A_i$, of the segments and the detuning of the pulse are optimized to minimize the measurement-induced dephasing due to residual photon number in the resonator after the given readout duration, $\tau = \sum_i\tau_i$, with respect to the signal-to-noise ratio. The \textit{readout power} of these pulses is labeled by the expected steady-state photon number with respect to the second segment of the pulse, $A_2$.
In this optimization, we thus restrict the instantaneous photon number of the resonator below this photon number at all times.

\begin{figure}[t]
    \centering
    \includegraphics[width=0.45\textwidth]{ 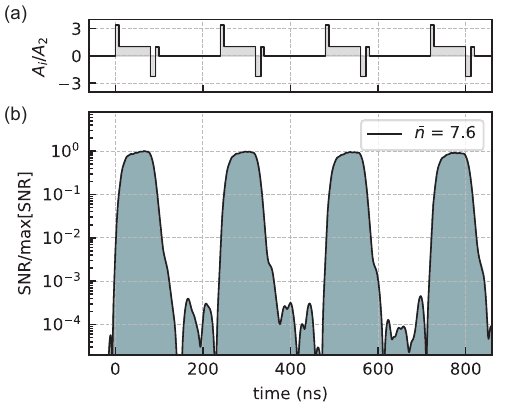}
    \caption{Repeated readout with active resonator reset pulse for $\bar{n} = 7.6$ in Exp 2. (a) The $100$ ns readout pulse including the resonator reset segments. Two successive pulses are separated by $140$ ns during which the pi pulse is played. (b) Measured $|\alpha_g(t)-\alpha_e(t)|^2$, normalized by its maximum value, showing the emptying of the readout resonator below $1\%$ by the end of the readout pulse.}
    \label{fig:active_reset}
\end{figure}

Experimentally, the quantity to be minimized is thus $\Gamma_{\rm residual}/\rm SNR$. Therefore, we can formulate the optimization problem with a cost function:
\begin{equation}
    \label{eq:cost_function_pulse}
    \texttt{Minimize}\quad\vartheta(A_i, \tau_i, \Delta)=\frac{\int_\tau^{10\tau}{|\alpha_g(t)-\alpha_e(t)|^2dt}}{\int_0^\tau{|\alpha_g(t)-\alpha_e(t)|^2dt}} \quad \texttt{such that}\quad  \max\left[|\alpha_g(t)|^2, |\alpha_e(t)|^2\right]\leq\bar{n}_{\rm max}
\end{equation}
We have optimized the readout pulse for both Exp 1 (cooldown I) and Exp 2 (cooldown III) at two different readout powers, corresponding to photon numbers of $4.9$ and $7.6$, respectively. The segment lengths were chosen to be integral multiples of $4$ due to hardware constraints. The lengths and amplitudes of the segments for each of the optimized pulses are shown in Table~\ref{tab:ro_pulse_parameter}
We illustrate the resulting signal trajectory $|\alpha_g(t)-\alpha_e(t)|^2$ as a function of time, for repeated application of the readout pulse in Fig.~\ref{fig:active_reset}. Note that this calibration experiment is performed without a quantum-limited amplifier to avoid slowing down of the output pulse due to the limited gain-bandwidth~\cite{_hazra_dai_2025} of a non-impedance-engineered amplifier.

\begin{figure}[b]
    \centering
    \includegraphics[width=0.95\textwidth]{ 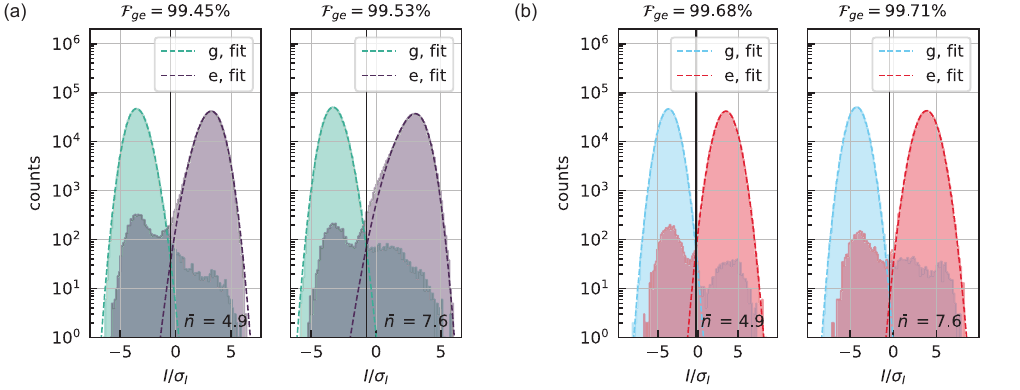}
    \caption{Readout histograms constructed from $700\,000$ measurement shots each, for (a) Exp 1 and (b) Exp 2, with two different readout powers, $\bar n = 4.9$ and (b) $\bar n = 7.6$, respectively. The readout fidelities in Exp 1 are $\mathcal{F}_{ge}=99.45\%$ and $\mathcal{F}_{ge}=99.53\%$, respectively for $\bar n = 4.9$ and $\bar n = 7.6$. In Exp 2, the fidelities in the two readout power settings are $\mathcal{F}_{ge}=99.68\%$ and $\mathcal{F}_{ge}=99.71\%$, respectively. The demarcation  (vertical black solid line) for state assignment is computed from fitting a skewed Gaussian function (Exp 1) and a standard Gaussian function (Exp 2) to each of the distributions corresponding to ground and excited states.}
    \label{fig:readout_hist}
\end{figure}

\subsection {Single-shot readout performance}
We construct the single-shot readout histograms from the measurement records of the first readout pulse of the repeated readout sequence. We pick $700\,000$ random uniform samples from the ensemble to construct the readout histograms in each case. These four pairs of histograms for each of the two readout powers in Exp 1 and  Exp 2 are shown in Fig.~\ref{fig:readout_hist}(a) and (b), respectively. In each plot, we have normalized the $x$-axis by the average standard deviation, $\sigma_I=(\sigma_I^g+\sigma_I^e)/2$, of the fitted distributions for the ground and the excited states.

In Exp 1, the readout is affected by leakage into higher excited states, as revealed by the leakage benchmarking experiment. This effect is also observed in the readout histograms as the distribution of the excited state signals deviates from a Gaussian shape. We fitted the distribution with a skewed Gaussian (dashed lines) in this case, as shown in Fig.~\ref{fig:readout_hist}(a). However, in Exp 2, as this leakage becomes negligible, the distributions recover their Gaussian shapes, and we fit the distribution without any skewness parameter, as shown in Fig.~\ref{fig:readout_hist}(b).

In each readout setting, the demarcation threshold, shown by the vertical black solid line, for the binary readout outcome is determined by the intersection point of the two fitted distributions. We then quantify the readout fidelity in each case with respect to this demarcation threshold. The readout fidelities in Exp 1 are $\mathcal{F}_{ge}=99.45\%$ and $\mathcal{F}_{ge}=99.53\%$, respectively for $\bar n = 4.9$ and $\bar n = 7.6$. In Exp 2, the fidelities in the two readout power settings are $\mathcal{F}_{ge}=99.68\%$ and $\mathcal{F}_{ge}=99.71\%$, respectively. The readout photon number is calibrated by performing an ac Stark shift measurement of the qubit~\cite{_hazra_dai_2025}.

\
\section{Spectroscopy of drive-induced unwanted state transitions}

We perform a \emph{time-resolved pump-probe spectroscopy} experiment to map the landscape of drive-induced unwanted state transitions (DUST) as a function of drive frequency and drive power, as visualized in the main text Fig.~3.
In this appendix, we present experimental results of a pump frequency sweep spanning over $4$ GHz on the dimon device, and discuss the principal families of observed transitions, which we classify into three
mechanisms~\cite{_dai_hazra_weiss_2025}: (1) Resonant exchange with two-level system (TLS) defects, (2) Intrinsic multiphoton excitations, and (3) Inelastic scattering involving external modes.

\subsection{Pump-probe spectroscopy experiment setup and calibration}
\label{app:dust:method}
The dimon circuit energy levels and readout resonator frequency in this experiment are characterized as shown in Fig.~\ref{fig:system_and_setup}(c). 
The pulse sequence for the experiment is similar to Ref.~\cite{_dai_hazra_weiss_2025}. 
We prepare the qubit in either $\ket{g}\equiv \ket{0_q 0_m}$ or $\ket{e}\equiv\ket{1_q 0_m}$, apply a $1\,\mu\mathrm{s}$ stimulation pulse of variable drive frequency $\omega_d$ and amplitude, wait $400$~ns to ensure the readout cavity is empty, and finally perform a single-shot multi-state readout of the dimon. 
The observed transition probabilities are shown in Fig.~\ref{fig:dust_sweep}. 
For this experiment, the stimulation drive is applied from the asymmetrically coupled port $P_{\rm asym}$, therefore relaxing the selection rule for transitions in the dimon circuit. 

The stimulation drive is synthesized and calibrated in the same manner as explained in Ref. \cite{_dai_hazra_weiss_2025} Appendix B.1. 
We calibrate the drive power by the resulting ac-Stark shift on the qubit mode $\Delta_q^{\rm ac}$. 
At each drive frequency, we record the slope of $\Delta_q^{\rm ac}$ versus the applied drive power. We then scale the power independently at each frequency according to this slope, so that $|\Delta_q^{\rm ac}/\alpha_q|$ ranges from $0$ to $1/2$, where $\alpha_q/2\pi = -182.2$ MHz is the anharmonicity of the qubit mode. 

In sweeping across the full $4.8$-$9$~GHz frequency range, different amplifier and band-pass filter configurations are employed to achieve the desired power levels while suppressing spurious tones. 
The gray hatched regions in Fig.~\ref{fig:dust_sweep} mark drive-frequency windows in which the room-temperature electronics cannot deliver enough power to reach the intended Stark shift while remaining in their linear regime. 
We also exclude a window around $\omega_d \approx \omega_q$ where the drive is expected to activate Rabi oscillations for $\ket{g}\leftrightarrow\ket{e}$ and $\ket{e}\leftrightarrow\ket{f}$ transitions. 
Driving above the qubit transition frequency, we observe oscillation fringes due to the detuned Rabi drive, which reflects the asymmetric coupling of the drive to the dimon.

\subsection{Resonant exchange with TLS defects}

\begin{figure*}[b]
    \centering
    \includegraphics[width=\textwidth]{ 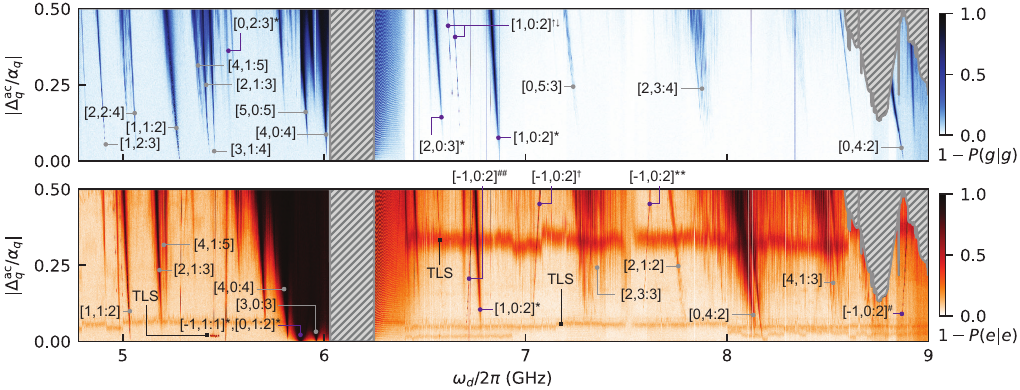}
    \caption{Measured drive-induced unwanted state transitions in dimon through the asymmetric port. Time-resolved pump-probe spectroscopy is performed with pump frequencies sweeping between $4.8$ GHz and $9$ GHz. The pump power is calibrated at each frequency via the induced ac Stark shift on the qubit mode. The extent of the pump power is chosen such that the maximum ac Stark shift equals half the anharmonicity of the qubit mode. 
    Identified transitions are labeled by $\left[q,m:d\right]$, denoting $d$ drive quanta exciting the qubit and the mediator mode by $q$ and $m$ quanta, respectively. The intrinsic transitions are identified through Floquet simulations. The transitions arising from inelastic scattering processes are denoted by $\left[q,m:d\right]^{[\quad]}$ where the superscript denotes the parasitic mode involved in the process. We have verified the presence of these parasitic modes through a finite element simulation of the package, identified at
    $\omega^{*}/2\pi = \omega_r/2\pi = 7.513$ GHz, $\omega^{\mbox{\scriptsize{\texttt{\#\#}}}}/2\pi = 19.63$ GHz, $\omega^\dagger/2\pi = 20.28$ GHz, $\omega^{**}/2\pi = 21.38$ GHz,  and $\omega^{\mbox{\scriptsize{\texttt{\#}}}}/2\pi = 23.95$ GHz. The transition, labeled $\left[1,0:2\right]^{\uparrow\downarrow}$, arises from inelastic scattering with a fluctuating two-level system (TLS) defect. In addition, the transitions from the excited state also feature quasi-horizontal lines representing resonant exchange with two-level system (TLS) defects.
    }
    \label{fig:dust_sweep}
\end{figure*}

We first focus on the quasi-horizontal features visible only in the lower panel of Fig.~\ref{fig:dust_sweep}, i.e., with initialization to $\ket{e}$. Through readout histograms, we verify that for these resonances, the qubit transitions into $\ket{g}$. 
Such features are signatures of a resonant energy exchange between the qubit mode and a discrete dissipative bath of microscopic defects generally known as two-level systems (TLS)~\cite{_Martinis2005_TLS, _Klimov2018, _chen2024_tls, _lisenfeld2019_tls, _cole2010_TLS, _burnett2019_TLS, _spiecker2023}. 

These transitions are activated whenever $\tilde\omega_q=\omega_{\rm TLS}$, where $\tilde\omega_q = \omega_q + \Delta_q^{\rm ac}$ is the dressed qubit frequency that depends only on the ac Stark shift, not on the drive frequency. Therefore, such transitions appear as horizontal lines in the $(\omega_d,\Delta_q^{\rm ac})$ plane. 
These lines show drifting or switching behaviors throughout the sweep, manifesting the temporal instability of the TLS frequencies~\cite{_Klimov2018}. 
The Stark shift at which a certain TLS shows up is observed to drift by tens of MHz over the course of hours, and can undergo telegraphic jumps between two well-defined values. 
Because TLS defects are microscopic, they do not respect the device symmetry and can couple linearly to the quadrupolar qubit mode and resonantly exchange energy.

\subsection{Intrinsic multiphoton excitations}

We next discuss the features in Fig.~\ref{fig:dust_sweep} that appear as sloped lines whose position depends on both $\omega_d$ and $\Delta_q^{\rm ac}$. 
The Floquet steady-state simulation, as shown in Fig.~2(d), can predict those arising from multiphoton resonances intrinsic to the dimon Hamiltonian \ref{eq:harmonic_hamiltonian}. 

The dimon circuit, in which both the qubit mode and the mediator mode are transmon-like degrees of freedom, is a minimal multimodal system. During this experiment, the mediator mode in our dimon device has frequency $\omega_m/2\pi = 4.605$ GHz and anharmonicity $\alpha_m/2\pi = -102$ MHz, both on the same orders of magnitude as those of the qubit mode, with a cross-Kerr $\chi_{qm} = -273$ MHz between the two modes. Note that this device \textit{qualitatively} differs from Ref.~\cite{_mori2025_nonlin}, in which the qubit mode is transmon-like, and the mediator (auxiliary) mode is generalized flux qubit-like~\cite{_yan2020_gen_fluxq}, operated at zero external flux.

In our case, the resulting energy level spectrum is denser than that of a single transmon, as shown in Fig.~\ref{fig:system_and_setup}(c), over the energy range of relevance when driving at a comparable frequency.
Both self- and cross-mode nonlinearity of the system result in significant matrix elements for the allowed multiphoton transitions. The broadband spectrum of multiphoton resonances has been previously demonstrated in a transmon through a pump-probe spectroscopy technique~\cite{_dai_hazra_weiss_2025}. Here we extend this technique to characterize all the multiphoton processes present in the multimodal Josephson circuit for a frequency range of interest spanning over $4$ GHz.

Intuitively, both the qubit mode and the mediator mode now have all possible multiphoton resonances as a transmon. In addition to those, there are resonances involving joint excitation of the two modes. 
Here we use the shorthand notation $[q,m{:}d]$ to label a transition that absorbs $d$ drive photons to excite the qubit mode by $q$ quanta and the mediator mode by $m$ quanta. 
Naively, a quadratic expansion of the dimon Hamiltonian approximates the energy level $\ket{q_q m_m}$ at: 
\begin{equation}
  \omega_{qm} \approx q\,\omega_q + m\,\omega_m + \frac{q(q-1)}{2}\alpha_q + \frac{m(m-1)}{2}\alpha_m + q\,m\,\chi_{qm}, 
  \label{eq:dimon_levels}
\end{equation}
providing a qualitative guide to narrow down candidate $[q,m{:}d]$ resonances within a drive frequency range. 

To identify each feature quantitatively, we perform a Floquet steady-state simulation~\cite{_floquet2024}.
The parameters in the dimon Hamiltonian are fitted independently from the characterized level spectrum shown in Fig.~\ref{fig:system_and_setup}(c) to be $E_J = 16.52$ GHz, $E_{C_J}=0.3221$ GHz, and $E_{C_s}=0.775$ GHz. 
The higher levels deviate from the prediction of this Hamiltonian, which can be accounted for by the higher harmonics of the potential~\cite{_willsch2024_Harmonics,_kim2026harmonics} and level repulsion due to hybridization with higher cavity modes. 

We have compared all the observed features with the Floquet simulation in the $(\omega_d, \Delta_q^{\rm ac})$ map. 
For each candidate transition, we refer to a branch analysis to identify the non-computational state involved in the resonance. 
Multiple resonances can further hybridize with each other, complicating the identification of the features. 
We locate, to the best of our ability, the lowest-order resonance process associated with each predicted feature, as labeled in Fig.~\ref{fig:dust_sweep}. 
The same resonance starting from $\ket{e}$ always occurs at a systematically lower drive frequency than that from $\ket{g}$ due to the negative anharmonicity and cross-Kerr interaction. 
Similar to the offset charge dependence of DUST in a transmon~\cite{_Fechant2025}, we observe \textit{smearing} of some transition features due to the resonances involving charge-sensitive final states. 

As expected, the majority of the features are resonances involving joint excitation of the two modes. Remarkably, when the drive frequency is near the mean of the two mode frequencies, $\omega_m/2\pi = 4.605$ GHz and $\omega_q/2\pi = 6.231$ GHz, the spectrum gets crowded with a large number of multiphoton processes. 
There are many low-order resonances, such as $[1,1{:}2]$, $[1,2{:}3]$, $[2,1{:}3]$, lying within a few hundred MHz from this center frequency, making this frequency range inaccessible for engineering any parametric process.

\subsection{Inelastic scattering involving external modes}

The remaining sloped features in Fig.~\ref{fig:dust_sweep} (those not predicted by the Floquet simulation) are inelastic scattering processes involving parasitic modes. 
Inelastic processes can, in principle, scatter photons into the continuum~\cite{_connolly_2025} along with transitions in the Josephson circuit. 
The parasitic modes create a high density of states in the electromagnetic environment, enhancing the inelastic processes that hit a resonance with output photons at specific frequencies. 
As a result, the transitions of this class also appear as resonant features in the $(\omega_d, \Delta_q^{\rm ac})$ plane. 

We denote such processes by appending a superscript symbol to the $[q,m{:}d]$ label, with the superscript indicating which parasitic mode is involved. 
For instance, we label all processes involving the readout cavity mode $\omega_r/2\pi = 7.513$ GHz with a superscript $*$. 
Take $[1,0{:}2]^{*}$ as an example: two drive photons excite the qubit mode by one quantum and emit one photon into the readout cavity, with a 4-wave-mixing resonance condition: 
$2\omega_d \;=\; \tilde\omega_q + \omega_r.$
To leading order, it results in a line intercepting the x-axis at $(\omega_q + \omega_r)/2 = 2\pi \times 6.87$~GHz, with a slope of $1/2$ in the $(\omega_d, \Delta_q^{\rm ac})$ plane. 
The corresponding process from the excited state, $\ket{1_q 0_m} \to \ket{2_q 0_m}$ requires a drive at $(\tilde\omega_{q} + \alpha_q + \omega_r)/2$ instead. As a result, it appears systematically shifted to a lower drive frequency by around $90$~MHz. 
Though the quadrupolar nature of the qubit mode is supposed to suppress such a process, the remaining asymmetry is enough to activate this process. 
Using a similar frequency matching criteria, we are also able to identify 6-wave-mixing processes involving the readout mode $[2,0{:}3]^{*}$ and $[0,2{:}3]^{*}$, that excites the qubit mode and mediator mode by 2 quanta respectively. 

The lines with opposite slope, labeled $[-1,0{:}2]^{[\,\cdot\,]}$, are processes in which the qubit relaxes from $\ket{e}$ to $\ket{g}$ through a 4-wave-mixing interaction that takes two drive photons to convert the qubit excitation into a quantum of a higher-frequency parasitic mode. 
We locate the frequencies of the relevant modes, based on 4-wave-mixing theory, to be: 
\begin{align}
\omega^{\mbox{\scriptsize{\texttt{\#\#}}}}/2\pi &= 19.63\;\mathrm{GHz},\quad
  \omega^{\dagger}/2\pi   = 20.28\;\mathrm{GHz}, \nonumber\\
  \omega^{**}/2\pi   &= 21.38\;\mathrm{GHz},\quad
\omega^{\mbox{\scriptsize{\texttt{\#}}}}/2\pi   = 23.95\;\mathrm{GHz}.
  \label{eq:parasitic_freqs}
\end{align}

\begin{figure}[t]
    \centering
    \includegraphics[width=0.95\textwidth]{ 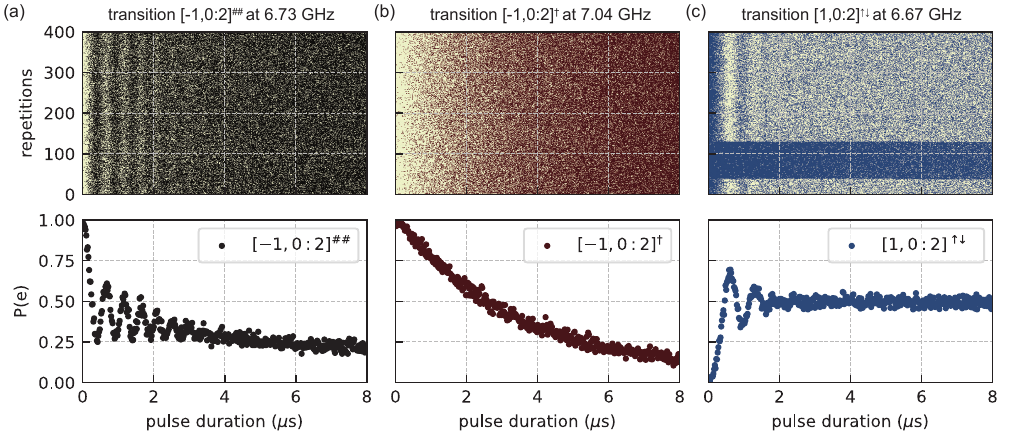}
    \caption{Measured time domain dynamics on three distinct classes of inelastic scattering processes: (a) A coherent ($g_{\rm eff}>\Gamma$) exchange with a parasitic mode, (b) An incoherent ($g_{\rm eff}<\Gamma$) transfer to a parasitic mode, and (c) A coherent ($g_{\rm eff}>\Gamma$) parametric interaction with a temporally switching mode. $g_{\rm eff}$ measures the effective transition rate between the two involved levels, and $\Gamma$ corresponds to the decoherence rate of the final state. Each of the points in the top panel represents the single-shot binary outcome, with bright points denoting an inferred state $|e\rangle$, and the dark colored points represent an inferred state $|g\rangle$. In fig (c), the blue band, where the dynamics is absent, suggests that the process involves a mode that is switching in time. The bottom panel displays the ensemble-averaged dynamics of the driven system for these three cases. 
    }
    \label{fig:time_domain}
\end{figure}

The pump-probe spectroscopy of Fig.~\ref{fig:dust_sweep} identifies \emph{where} a transition becomes resonant in the $(\omega_d, \Delta_q^{\rm ac})$ plane. 
To further characterize these transitions, we investigate the time dynamics of three representative features that are associated with inelastic scattering processes. 
We lock the drive on resonance and sweep the pulse duration to record the final state after the drive pulse. 
The resulting dynamics, as shown in Fig.~\ref{fig:time_domain}, depend on the ratio between the drive-activated effective coupling strength $g_{\rm eff}$ and the decoherence rate $\Gamma$ of the external mode. 
We show three qualitatively distinct regimes: 
Transition $[-1,0{:}2]^{\mbox{\scriptsize{\texttt{\#\#}}}}$ involves a high-$Q$ electromagnetic package mode at $\omega^{\mbox{\scriptsize{\texttt{\#\#}}}}$, with the coherent-exchange relation $g_{\rm eff} > \Gamma$. 
The average population oscillates over time. 
Transition $[-1,0{:}2]^{\dagger}$ places $\omega^{\dagger}$ in the overdamped limit $\Gamma > g_{\rm eff}$, where the qubit excitation is irreversibly absorbed at a Purcell-like rate $\kappa_{\rm leak} \approx 4 g_{\rm eff}^2 / \Gamma$. 

There is one transition labeled $[1,0{:}2]^{\updownarrows}$, which according to the frequency-matching condition based on 4-wave-mixing theory should involve a spurious mode at $\omega^{\updownarrows}/2\pi \approx 7.2$ GHz. 
When looking at the time domain results shown in Fig.~\ref{fig:time_domain}(c), we see oscillating dynamics that switch on and off over time. 
In the unaveraged shots data (upper panel), there is a block of shots for which the oscillating transition is not activated. 
The turn-on and off of the transitions appear abrupt across the repetition of shots. 
This feature reflects that the frequency of the $\omega^{\updownarrows}/2\pi \approx 7.2$ GHz spurious mode involved in this transition is unstable, suggesting its nature as a TLS defect in the system. 

Through finite element simulation, we identify the relatively low Q modes $ \omega^{**}$, $\omega^{\mbox{\scriptsize{\texttt{\#}}}}$, and $\omega^{\dagger}$, within $1.5\%$ accuracy from the experimentally predicted value. 
Finally, we expect a relatively higher Q mode near $\omega^{\mbox{\scriptsize{\texttt{\#}}\texttt{\#}}}$, which is associated with the coherent oscillations observed in~\ref{fig:time_domain}(a). 
We have not observed any temporal switching of this mode within a $\sim1$ hour timescale. Therefore, we conclude that this transition is associated with either a temporally stable coherent TLS or a high Q package mode.

\subsection{Finite element simulation of the qubit's linear RF environment}
As discussed in the previous section, several spurious transitions result from inelastic scattering with a parasitic RF mode in the package (in this case, a rectangular cavity and waveguide coupler system). It is thus crucial to identify and suppress these modes beforehand to reduce DUST arising from these modes.
In this section, we discuss how we simulate the linear RF environment, $Y_{\rm env}[\omega]$ seen by the qubit mode embedded in the multimodal circuit. We follow the black-box quantization approach~\cite{_nigg2012_bbq} to derive the effective qubit environment from a finite element High Frequency Simulation Software (HFSS) simulation. 

We perform a driven modal simulation in Ansys HFSS, in which the junctions are replaced by lumped ports, $P_1$ and $P_2$, and the waveguide port is terminated with a matched $50~\Omega$ load. In this simulation, we displace the chip by $0.5$ mm from the center of the cavity to account for any chip misalignment in realistic setups.
We first extract the admittance matrix of this two-port network, and then shunt the two ports $P_1$ and $P_2$ with lumped inductors $L_1$ and $L_2$, respectively, which correspond to the linearized Josephson inductance. The asymmetry of the pair of junctions is also modeled through the values of these inductors, $L_{1,2}=L(1\pm \epsilon/2)$, where the values are obtained from the frequencies of the qubit and the mediator mode. 
\begin{equation}
    \mathbf{Y}_{\rm node}= 
    \begin{bmatrix}
        Y_{11}+\left(j\omega L_1\right)^{-1} & Y_{12} \\
        Y_{21} & Y_{22}+\left(j\omega L_2\right)^{-1} 
    \end{bmatrix}
\end{equation}

\begin{figure*}[b]
    \centering
    \includegraphics[width=0.95\textwidth]{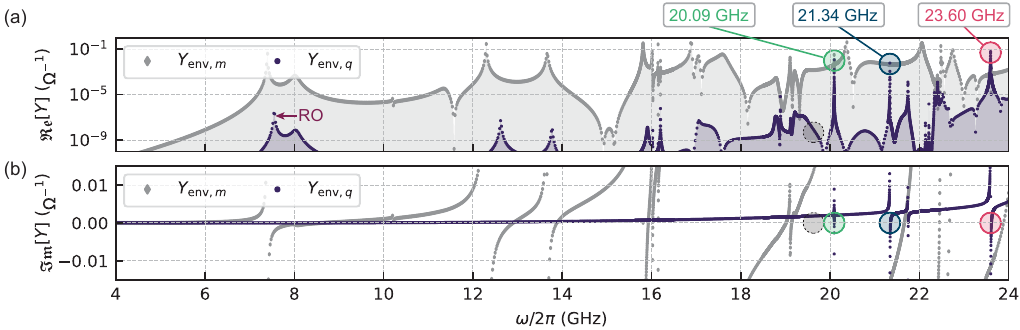}
    \caption{Cavity modes from driven modal simulation in HFSS with interpolated simulation. The plots are generated with a discrete simulation, in which the full electromagnetic environment of the meshed system is solved at every frequency. The simulated cavity modes (indicated by solid colored circles) agree within $\lesssim 1.5\%$ accuracy in frequency to those inferred from the DUST spectroscopy experiment as shown in Eq.~\ref{eq:parasitic_freqs}. Note that a fourth high-Q mode at $19.63$ GHz is expected from the spectroscopy experiment (marked by the gray dashed circle), which could not be resolved in the driven modal simulation due to sampling limitations.
    }
    \label{fig:cavity_modes}
\end{figure*}

Then we transform the ports from node excitations basis to mode excitations basis through eigen-decomposition, $\mathbf{Y}_{\rm node} = U\mathbf{Y}_{\rm mode} U^{-1}$, leading to common and differential mode excitations~\cite{_pozar2012microwave}. The modal coordinates are $\mathbf V_{\rm mode} = U^{-1}\mathbf V_{\rm node}$ and $\mathbf I_{\rm mode} = U^{-1}\mathbf I_{\rm node}$. In this basis, the admittance matrix is diagonal:
\begin{equation}
    \mathbf Y_{\rm mode}= 
    \begin{bmatrix}
        Y_{{\rm in}, m} & 0 \\
        0 & Y_{{\rm in}, q}
    \end{bmatrix},
\end{equation}
where $Y_{{\rm in}, q}\left[\omega\right]$ $Y_{{\rm in}, m}\left[\omega\right]$ are the input admittances seen from differential (quadrupolar or qubit excitation) and common (dipolar or mediator excitation) ports, respectively. 
This computed input admittance also includes the qubit and the mediator modes. This self-contribution is then subtracted to derive the environmental admittance seen by the qubit or mediator mode:
\begin{align}
    Y_{{\rm env}, q}\left[\omega\right]=Y_{{\rm in}, q}\left[\omega\right]-j\omega C_q-\frac{1}{j\omega L_q}, \\
    Y_{{\rm env}, m}\left[\omega\right]=Y_{{\rm in}, m}\left[\omega\right]-j\omega C_m-\frac{1}{j\omega L_m},
\end{align}
where $C_q$ ($C_m$) and $L_q$ ($L_m$) are the effective capacitance and inductance of the qubit (mediator) mode, obtained from fitting ${\rm Im}Y_{\rm mode}[\omega]$ near the two associated mode frequencies. 

In Fig.~\ref{fig:cavity_modes}(a), we plot the simulated environment admittance $Y_{{\rm env}, m}$ and $Y_{{\rm env}, q}$, seen from the excitation port of the mediator mode and the qubit mode, respectively. The linear couplings of the qubit to the relatively lower-frequency package modes (such as the readout mode) are minimized by design, protecting the qubit from Purcell decay. At these frequencies, the input admittance seen by the qubit mode is also suppressed. However, at higher frequencies, the qubit strongly couples to several package modes, making it susceptible to parasitic two-mode parametric conversion and squeezing processes, as explained in the previous section. We identify three of the four parasitic modes mentioned in Eq.~\ref{eq:parasitic_freqs}. Since the remaining mode at $\omega^{\mbox{\scriptsize{\texttt{\#\#}}}}$ has a high quality factor, we could not resolve it in the frequency sweep performed in HFSS. These identified modes are shown by green, blue, and pink circles in Fig.~\ref{fig:cavity_modes}. The identified modes are within $1.5\%$ agreement with the experimentally measured ones. Since the high-frequency modes are sensitive to the package dimensions, the mismatch could be attributed to a deviation between the experimental package and the modeled one.

Note that these results are from a linearized RF model of the system to identify the modes involved in the inelastic scattering processes observed in the DUST spectroscopy experiment. In principle, inelastic scattering can happen even through intrinsic nonlinear couplings. An example of such a process is the drive exciting the qubit mode by two quanta and depositing the remaining energy into one of these parasitic modes. Such processes will not be captured in this model. The spectrum is progressively denser at higher frequencies as the mode index grows. The frequencies of these modes are thus sensitive to the exact dimensions of the simulated RF enclosure. At these higher frequencies, the simulated result will, in general, deviate further from the experimentally measured package. Further, this model does not account for the coupling of the system to the asymmetric drive port, through which the spectroscopy drive is sent.

\section {Readout-induced leakage benchmarking}
\label{rilb_2}

We first briefly compute the leakage state population after $m$ cycles and then derive an expression for the measured correlation, accounting for experimental non-idealities.

The Hilbert space is divided into a $ d_1=2$-dimensional computational subspace and a $ d_2$-dimensional non-computational subspace. The total leakage population is given by  $\mathfrak{L}(\rho) = {\rm{Tr}}[\mathbb{1}_2\rho]$, where $\mathbb{1}_2$ is the projector onto the non-computational subspace~\cite{_wood_lrb_2018}.
The leakage error channel is described as a completely positive trace-preserving (CPTP) map, $\mathcal{E}_L$ in the $(d_1+d_2)$-dimensional space. When this leakage error channel acts on a state, $\psi_1
$, in the computational subspace, the state may transition into the non-computational space with a total probability: $p_l = \mathfrak{L}\left(\mathcal{E}_L\left[|\psi_1\rangle\langle\psi_1|\right]\right)$. Similarly, if this error channel acts on a state outside of the computational subspace (or a leakage state), the system may transition back into the computational subspace with a probability $p_s = 1-\mathfrak{L}\left(\mathcal{E}_L\left[|\psi_2\rangle\langle\psi_2|\right]\right)$. We call this process ``seepage''.

In our first model, we consider only the average leakage and seepage rates, $L_{\uparrow}$ and  $L_{\downarrow}$, defined for transitions from the computational subspace to the leakage states and vice versa, without resolving the specific initial or final states.
Within this simplified framework, the average leakage and seepage rates are defined as follows: 
\begin{align}
    L_\uparrow\left(\mathcal{E}_L\right) = \int d\psi_1\mathfrak{L}\left(\mathcal{E}_L\left[|\psi_1\rangle\langle\psi_1|\right]\right)= \mathfrak{L}\left(\mathcal{E}_L\left(\frac{\mathbb{1}_1}{d_1}\right)\right),
\label{eq:dle_avg}
\end{align}

\begin{align}
    L_\downarrow\left(\mathcal{E}_L\right) = 1-\int d\psi_l\mathfrak{L}\left(\mathcal{E}_L\left[|\psi_l\rangle\langle\psi_l|\right]\right)=1-\mathfrak{L}\left(\mathcal{E}_L\left(\frac{\mathbb{1}_2}{d_2}\right)\right).
\label{eq:dle_avg2}
\end{align}

Assuming no coherence between the computational and non-computational subspaces, the superoperator for the CPTP map $\mathcal{E}_L$ is expressed in the basis, $\|\mathbb{1}_1\rrangle \equiv (1, 0)$ and $\|\mathbb{1}_2\rrangle  \equiv (0, 1)$ as a $2\times2$ matrix:

\begin{align}
    \mathcal{S}_{\mathcal{E}_L}=
    \begin{pmatrix}
        1-L_\uparrow & L_\downarrow\\
        L_\uparrow & 1-L_\downarrow
    \end{pmatrix}. 
\label{eq:leakage_model}
\end{align}

Next, the leakage population after $m$ applications of the cycle is given by:
\begin{align}
    \mathfrak{L}\left(\mathcal{E}_L^m[\rho]\right) = {\rm{Tr}}\left[\mathbb{1}_2\mathcal{E}_L^m\left[\rho\right]\right] = {\rm{Tr}}\left[\rho\left(\mathcal{E}_L^\dagger\right)^m[\mathbb{1}_2]\right].
\label{eq:err_exp}
\end{align}
The superoperator for the $m$-th power of the Hermitian conjugate of the leakage error channel is given by:
\begin{align}
    \mathcal{S}_{\mathcal{E}_L^\dagger}^m=\frac{1}{L}
    \begin{pmatrix}
        L_\downarrow & 
        L_\uparrow \\
        L_\downarrow  & 
        L_\uparrow
    \end{pmatrix} +
    \frac{(1-L)^m}{L}
    \begin{pmatrix}
        L_\uparrow & 
        -L_\uparrow \\
        -L_\downarrow  & 
        L_\downarrow
    \end{pmatrix}.
\label{eq:leakage_model_conjugate}
\end{align} 
We have introduced the notation $L = L_\uparrow +L_\downarrow$ to represent the total leakage-seepage rate. We apply this superoperator to the basis state $\|\mathbb{1}_2\rrangle$ to obtain:
\begin{align}
\begin{split} \mathcal{S}_{\mathcal{E}_L^\dagger}^m\|\mathbb{1}_2\rrangle &= 
   \frac{L_\uparrow}{L}
    (\|\mathbb{1}_1\rrangle+\|\mathbb{1}_2\rrangle) \\
    &-
    \frac{(1-L)^m}{L}
(L_\uparrow\|\mathbb{1}_1\rrangle-L_\downarrow\|\mathbb{1}_2\rrangle)
\end{split}
\label{eq:sup_op_exp}
\end{align}

Finally, using Eq.~\ref{eq:err_exp} and Eq.~\ref{eq:sup_op_exp}, we compute the total leakage population after $m$ applications of the detection cycle, when the system is initialized in the computational subspace, $\rho = \|\mathbb{1}_1\rrangle$ (See also Ref.~\cite{_marxer2026_iqm}):

\begin{align}
\begin{split}P_L:=\mathfrak{L}\left(\mathcal{E}_L^m[\mathbb{1}_1]\right) = \llangle\mathbb{1}_1\|\mathcal{S}_{\mathcal{E}_L^\dagger}^m\|\mathbb{1}_2\rrangle
    = \frac{L_\uparrow}{L}\left[1-(1-L)^m\right]
\end{split}
\label{eq:leakage_prpb}
\end{align}

\textbf{Measured average correlation:}
We now compute how the leakage population $P_L = \mathfrak{L}\left(\mathcal{E}_L^m[\rho]\right)$ affects the measured average correlation in the original RILB sequence. We assume that the readout and leakage events are instantaneous, and readout, Pauli, and leakage errors are modeled as independent random events. 
If the $m$-th applied input operation is $X$, the correlation is $(+1)$ (outcome flips) in the following mutually exclusive cases:
\begin{enumerate}
    \item The qubit has not leaked, and the following holds for the last two ($m$ and $m-1$) detection cycles: in each cycle, either no error occurs, or both an SNR error and a Pauli error occur simultaneously. The probability of this “true positive” event is
    \begin{align*}
    \begin{split}
    P_{\rm tp1} &= \left[(1-\epsilon_{\rm SNR})(1-\epsilon_\sigma)+\epsilon_{\rm SNR}\epsilon_\sigma\right]^2(1-P_L)\\
    &=\mathcal F_{ge}^2(1-P_L)
    \end{split}
    \end{align*}
    where we define the readout fidelity: $\mathcal F_{ge}:=(1-\epsilon_{\rm SNR})(1-\epsilon_\sigma)+\epsilon_{\rm SNR}\epsilon_\sigma$.
    \item The qubit has not leaked, and each of the two cycles has either an SNR error or a Pauli error (but not both). The probability of this “true positive” event is
    \begin{align*}
    \begin{split}
    P_{\rm tp2} &= \left[\epsilon_\sigma(1-\epsilon_{\rm SNR})+\epsilon_{\rm SNR}(1-\epsilon_\sigma)\right]^2(1-P_L)\\
    &=\left(1-\mathcal F_{ge}\right)^2(1-P_L)
    \end{split}
    \end{align*}
    \item The qubit has leaked into non-computational states, but the readout outcome falls on opposite sides of the threshold in the two successive readout cycles. Denoting the probability that it falls on the same side as $|0\rangle$ by $\alpha = P(0|\psi_L)$, the “false positive” probability is,
    $$
    P_{\rm fp}= P_L\left[\alpha(1-\alpha)+(1-\alpha)\alpha\right]=2P_L\alpha(1-\alpha)
    $$
    \end{enumerate}
    The total probability of obtaining a correlated outcome for an $X$ input is then:
    \begin{align}
    \begin{split}
    P_{\rm corr}^X(m)=& \left[\mathcal{F}_{ge}^2+(1-\mathcal{F}_{ge})^2\right](1-P_L(m))\\
    &+2P_L(m)\alpha(1-\alpha)
    \end{split}
    \end{align}

For an $I$ input at the $m$-th operation, the correlation is $(+1)$ (no flip) in the following mutually exclusive cases:
\begin{enumerate}
    \item The qubit has not leaked, and in the two most recent detection cycles ($m$ and $m-1$), either no readout or Pauli error occurs, or both errors occur simultaneously. The probability is
    $$
   P_{\rm tn1}=\mathcal F_{ge}^2(1-P_L)
    $$
    \item The qubit has not leaked, and each detection cycle has either a readout or a Pauli error (but not both). The probability is
    $$
   P_{\rm tn2}=\left(1-\mathcal F_{ge}\right)^2(1-P_L)
    $$
    \item The qubit has leaked into non-computational states, and the readout outcomes fall on the same side of the threshold in successive measurements. The “false negative” probability is:
    $$
    P_{\rm fn}=P_L\left[\alpha^2+(1-\alpha)^2\right]
    $$  
\end{enumerate}
Hence, the total probability of getting a correlated outcome for $I$ inputs is:
\begin{align}
\begin{split}
 P_{\rm corr}^I(m)=& \left[\mathcal{F}_{ge}^2+(1-\mathcal{F}_{ge})^2\right](1-P_L(m))\\
 &+P_L(m)\left[\alpha^2+(1-\alpha)^2\right] 
\end{split}
\end{align}

Since the input sequence is randomized, the $X$ and $I$ inputs are equally likely at any given instant, and the mean correlation is given by:
\begin{align*}
\begin{split}
    \langle\bar{\mathcal{C}}_m\rangle &= \frac{\left(2P_{\rm corr}^I-1\right)}{2}+\frac{\left(2P_{\rm corr}^X-1\right)}{2}
    = P_{\rm corr}^I+P_{\rm corr}^X-1\\
    &= 2\left[\mathcal{F}_{ge}^2+(1-\mathcal{F}_{ge})^2\right](1-P_L)+P_L\left[\alpha+(1-\alpha)\right]^2-1\\
    &= 2\left[\mathcal{F}_{ge}^2+(1-\mathcal{F}_{ge})^2\right](1-P_L)+P_L-1\\
    &= (1-P_L)(2\left[\mathcal{F}_{ge}^2+(1-\mathcal{F}_{ge})^2\right]-1)
\end{split}
\end{align*}

Substituting the expression for $P_L$ from Eq.~\ref{eq:leakage_prpb} and simplifying:
\begin{align}
\langle\bar{\mathcal{C}}_m\rangle =\frac{L_\downarrow+L_\uparrow(1-L)^m}{L}\left(2\mathcal{F}_{ge}-1\right)^2,
\end{align}

where $L= L_\downarrow+L_\uparrow$.
Note that randomizing the input sequence with $X$ and $I$ operations leads to cancellation of the term $\alpha$, rendering the benchmarking protocol insensitive to the readout outcome of leakage states on the IQ plane.

State initialization at the beginning of the sequence is performed via a slower readout, post-selecting on the $0$ outcome.
Consequently, leakage during the zeroth readout is expected to be negligible, and for the purposes of calculation, we treat it as leakage-free, reflecting only the SNR and Pauli errors $\epsilon_{\rm SNR}$ and $\epsilon_\sigma$ of the first readout of the sequence. 
The complete expression for the decay for this particular implementation of RILB is therefore:
\begin{align}
\begin{split}
    \langle\bar{\mathcal{C}}_{m=0}\rangle 
    &= \left(2\mathcal{F}_{ge}-1\right) \quad\\
    \langle\bar{\mathcal{C}}_{m>0}\rangle&= \frac{L_\downarrow+L_\uparrow(1-L)^m}{L}\left(2\mathcal{F}_{ge}-1\right)^2
\end{split}
\label{eq:rilb_trajectory}
\end{align}
Note that the described protocol yields only the average leakage rate and cannot distinguish leakage originating from \(|g\rangle\) versus \(|e\rangle\).
%%%%%%%%%%%%%%%%%%%%%%%%%%%%%%%%%%%%%%%%%%%%%%%%%%
%SM bbl

\end{document}